\documentclass[11pt]{article}
\pdfoutput=1
\usepackage{jheparxiv}

\usepackage{amsmath}
\usepackage{mathtools}
\usepackage{amssymb}
\usepackage{slashed}
\usepackage{color}
\usepackage{xcolor}
\usepackage{hyperref}
\hypersetup{colorlinks=true,citecolor=blue}
\usepackage{braket}
\usepackage{simplewick}
\usepackage{esint}
\usepackage{subcaption}
\usepackage[normalem]{ulem}
\allowdisplaybreaks

\newcommand{\Sprop}{S_{\textrm{in}}^{\textrm{c}}}
\newcommand{\sprop}{S^{\textrm{c}}}
\newcommand{\spropbar}{S^{\bar{\textrm{c}}}}
\newcommand{\Spropbar}{S^{\bar{\textrm{c}}}_{\textrm{in}}}
\newcommand{\llangle}{\langle\!\langle}
\newcommand{\rrangle}{\rangle\!\rangle}
\newcommand{\sgn}{\mathop{\mathrm{sgn}}}

\begin{document}

\title{In-in worldline formalism in pair creating fields}

\author[a,b]{Patrick Copinger}
\emailAdd{copinger@hiroshima-u.ac.jp}
\author[c,d]{and Shi Pu}
\emailAdd{shipu@ustc.edu.cn}

\affiliation[a]{International Institute for Sustainability with Knotted Chiral Meta Matter (WPI-SKCM${}^2$), Hiroshima University, 1-3-1 Kagamiyama, Higashi-Hiroshima, Hiroshima 739-8531, Japan}
\affiliation[b]{Centre for Mathematical Sciences, University of Plymouth, Plymouth, PL4 8AA, UK}
\affiliation[c]{Department of Modern Physics and Anhui Center for fundamental Sciences
(Theoretical Physics), University of Science and Technology of China,
Anhui 230026}
\affiliation[d]{Southern Center for Nuclear-Science Theory (SCNT), Institute of Modern
Physics, Chinese Academy of Sciences, Huizhou 516000, Guangdong Province,
China}

%\preprint{YITP-20-150}

\abstract{
An in-in framework under Schwinger pair creating fields in strong-field quantum electrodynamics is formulated using in-out propagators in coordinate space, that have first-quantized or worldline representation. The framework is derived to all orders in the background field coupling from both the Bogoliubov coefficient method and Schwinger-Keldysh closed-time path formalism. In-out matrix elements in pair creating fields are readily handled using first-quantized methods, and the approach we develop serves to facilitate the evaluation of in-in observables in pair creating backgrounds. We find that in-in augmentations to the in-out partition function and or propagator amount to the insertion of a non-local interaction term that sandwiches a function that receives contributions from singularities and critical points in complex Schwinger propertime. Furthermore, we show the resummation of the in-in partition function leading to vacuum non-persistence that en-route gives an exact first-quantized definition of creating $N$-pairs. 
}

\maketitle

%%%%%%%%%%%%%%%%%%%%%%%%%%%%%%%%%%%%%%%%%%%%%%%%%%%%%%%%%%%%%%%%%%%
\section{Introduction}

The Schwinger mechanism~\cite{Sauter:1931zz,Heisenberg:1936nmg,Schwinger:1951nm,Gelis:2015kya}, a vacuum instability against the production of particle anti-particle pairs in a strong electric field, is one of the most fundamental non-perturbative phenomena in quantum electrodynamics (QED). Recently, electromagnetic fields reaching magnitudes of $10^{17}-10^{18}$ Gauss, which are the strongest fields observed in the laboratory to date, have been generated in ultra-peripheral relativistic heavy-ion collisions \cite{Kharzeev:2007jp,Skokov:2009qp,Bzdak:2011yy,Voronyuk:2011jd,Deng:2012pc,Roy:2015coa,Pu:2016ayh,Roy:2015kma,Siddique:2019gqh}. These collisions provide a novel experimental platform for probing the physics of the strong-field regime. In ultra-peripheral collisions, the impact parameter exceeds twice the nuclear radius; the two nuclei are accelerated to near the speed of light but do not undergo direct hadronic contact. Consequently, strong interactions are highly suppressed, while QED effects are significantly enhanced by the large proton numbers of the nuclei. Recent observations in these collisions have confirmed the creation of dielectron pairs via the interaction of quasi-real photons (see Refs.~\cite{STAR:2018ldd,STAR:2019wlg, STAR:2004bzo, Zha:2018tlq, Klein:2018fmp, Li:2019yzy, Wang:2021kxm,Shi:2023nko} and references therein). While many studies have calculated the cross-sections for dilepton photoproduction using perturbative QED, particle production directly from the strong background field remains less explored. Investigating pair production in such strong-field environments necessitates a real-time formalism.

The evaluation of real-time observables in pair creating background field demands a framework in which expectation values rather than transition amplitudes can be found. Both the in-in~\cite{BakshiMahanthappa1963a,BakshiMahanthappa1963b} and Schwinger-Keldysh (SK) closed-time path~\cite{Schwinger:1960qe,Keldysh:1964ud} formalisms supply just that: by evolving operators or a doubled time contour one can compute real-time correlation functions that incorporate vacuum polarization and particle production. The in-in (and equivalently SK) formalism is particularly well suited for the Schwinger mechanism, where an in-out (S-matrix) machinery may miss important effects due to the mechanism in the evaluation of e.g. the mean current~\cite{fradkin1991quantum,Tanji:2008ku} or psuedoscalar condensate~\cite{Warringa:2012bq}. While the conventional in-out constructions yields vacuum persistence amplitudes and scattering observables, the in-in formalism naturally accommodates non-equilibrium physics, resummation, and finite-density effects~\cite{Berges:2004yj,Calzetta:2008iqa}, and is important in the study of strong-field QED~\cite{Fedotov:2022ely}.

However, one of our main theoretical tools for the study of the Schwinger effect, namely first-quantized methods including the worldline formalism~\cite{Feynman:1950ir,Feynman:1951gn,Strass1,ChrisRev,UsRep}, are defined with in-out matrix elements, giving us ease of access to just in-out derived observables, whereas in-in observables are inadequately addressed using the worldline formalism. The key merit of the formalism is in its all-orders in the (background) field coupling, an essential feature since the Schwinger effect is non-perturbative in the coupling. Thus, for example, the worldline formalism enables non-perturbative evaluations from a semi-classical standpoint--worldline instantons~\cite{Affleck:1981bma,Dunne:2005sx,Dunne:2006st,Dumlu:2011cc,Ilderton:2015lsa,Copinger:2020feb}--providing a means to treat \textit{any} background field, ones in which an eigendecomposition necessary for the evaluation of Bogoliubov coefficients might be impossible. And worldline techniques have been extended to non-Abelian d.o.f.~\cite{ChrisRev,103}, axial couplings~\cite{Bastianelli:2024vkp}, and even phase space~\cite{Migdal:1986pz,Copinger:2022jgg,Copinger:2022srm} to name a few, making the formalism widely applicable.  

A key example where usage of an in-in construction in a first quantization representation is beneficial lies with the evaluation of the chiral anomaly in electromagnetic parity-violating fields. A calculation of the divergence of the chiral current using a conventional in-out representation shows a cancellation of the anomalous term to the psuedoscalar condensate~\cite{Dittrich:2000zu}. However, non-conservation of the current--as predicted for the anomaly--is had with an in-in representation, indicating a central role of the Schwinger effect for the chiral anomaly~\cite{Copinger:2018ftr,Copinger:2020nyx}. However, a drawback in this example with homogeneous fields, as well as for other observables, is that a in-in worldline representation under a vacuum instability is only known for a handful of background fields--that treat the background field without recourse to perturbative analysis; see e.g.~\cite{fradkin1991quantum}.

In this work we develop a precise formulation of an in-in formalism in terms of in-out propagators that are expressible in first quantized form, or an in-in worldline representation, in a strong background field in QED. We accomplish this both from a Bogoliubov coefficient approach and from the SK closed-time path construction, where we show that the extension from the well-known in-out to in-in amounts to the insertion of a non-local interaction term serving to pick out the singularities due to a vacuum instability. We also go onto to show for the case of the in-in partition function a resummation structure in first-quantized form that leads to an exact definition for the $N$-pair creation rate, also in first-quantized form--analogous to the imaginary part of the effective action (the Euler-Heisenberg Lagrangian~\cite{Dunne:2004nc} for the case of homogeneous fields). 

In-in (and or equivalently SK) worldline formalisms have been been studied before: Ref.~\cite{fradkin1991quantum} notably derives in-in propagators using an eigendecomposition and Bogoliubov approach for select fields. Worldline SK theories where the non-trivial connection between path, a sum over states, has been represented by through a density matrix~\cite{Jalilian-Marian:1999uob,Mueller:2019gjj}. Our work aims to compliment existing studies, extending the formalism to encompass an all-orders background field construction for arbitrary field targeting the vacuum instability. Also, recently, worldline in-in/SK techniques have been actively applied to the problem of gravitational scattering amplitudes~\cite{Jakobsen:2022psy,Dlapa:2022lmu,Damgaard:2023vnx,Jakobsen:2021smu}. Also $N$-pair creation rates for the Schwinger effect have been studied in~\cite{Gelis:2006yv,Gelis:2006cr,Copinger:2024pai}, but a first-quantized form that we develop here has not yet been studied.

This manuscript is organized as follows: In Sec.~\ref{sec:basic} we lay out the essential framework of the Dirac operator notation and their various associated correlation functions. In Sec.~\ref{sec:in-in} we derive an in-in worldline formalism using both Bogoliubov coefficients and from the SK closed time-path. Next, in Sec.~\ref{sec:pairs} we treat the in-in generating functional in some depth, leading to a worldline representation for the creation of $N$-pairs. And last in Sec.~\ref{sec:conclusions} we discuss future works. 

We work in Minkowski spacetime with a mostly minus metric. Our QED covariant derivative reads $D_\mu=\partial_\mu+ieA_\mu$. And where a coincident limit appears we assume, e.g., for generic propagator, $S$: $\lim_{x\leftrightharpoons y}S(x,y)=(1/2)\bigl[\lim_{x\to y+\epsilon}+\lim_{x\to y-\epsilon}\bigr]S(x,y)$, which also serves to fix our Heaviside theta function so that $\theta(x-y)^2=\theta(x-y)$ and $\theta(x-y)\theta(y-x)=0$. For brevity Lorentz indices are left implicit where understood, and we also reserve boldface symbol notation, e.g., $\boldsymbol{x}$, for the three spatial dimension variables.

%%%%%%%%%%%%%%%%%%%%%%%%%%%%%%%%%%%%%%%%%%%%%%%%%%%%%%%%%%%%%%%%%%%
\section{Basic operator properties and in-(out/in) propagator definitions}
\label{sec:basic}

Let us first review some basic operator properties that allow us to define the in-out and in-in propagators. After which, we can show the various in-out propagators in their first-quantized form that will then later serve as the building blocks in an in-in construction.

In order to define our asymptotic vacuum states we first make use of the usual in and out operator formalism; we largely use notations as provided in Ref.~\cite{Gitman:1977ne,Fradkin:1981sc,fradkin1991quantum}. In a background Abelian electromagnetic field, the Dirac spinor field may be decomposed for any time as a sum over its eigenvectors with corresponding ``in'' or ``out'' creation and annihilation operators as
\begin{equation}
    \psi(x)=a_{n}^{\text{in}}\phi_{+n}^{\text{in}}(x)+b_{n}^{\text{in}\dagger}\phi_{-n}^{\text{in}}(x)=a_{n}^{\text{out}}\phi_{+n}^{\text{out}}(x)+b_{n}^{\text{out}\dagger}\phi_{-n}^{\text{out}}(x)\,,\label{eq:inout_field_def}
\end{equation}
where we keep an implicit summation for repeated eigenvalue indices, $n$.\footnote{A strategic aim of this work is to formulate the eigenmode resummed expressions in terms of first-quantized expressions, without actually calculating the eigenvalues. However, we demand that solutions to the Dirac equation in a background field admit an in/out decomposition as given in Eq.~\eqref{eq:inout_field_def}} $n$ for example can include momenta and Landau levels for homogeneous magnetic field backgrounds. The $\phi_{+n}^{\textrm{in/out}}$ or $\phi_{-n}^{\textrm{in/out}}$ are eigenvectors of the Dirac equation, $(i\slashed{D}_x-m)\phi_{\pm n}^{\textrm{in/out}}(x)=0$, with positive or negative energy at asymptotic (in/out) time with eigenvalue $n$, respectively. The eigenvectors satisfy some basic properties including orthonormality,
\begin{equation}
    \int\! d^{3}x\,\mathrm{tr}\bigl[\phi_{\pm n}^{\text{in}\dagger}(x)\phi_{\pm'm}^{\text{in}}(x)\bigr]=\int \!d^{3}x\,\mathrm{tr}\bigl[\phi_{\pm n}^{\text{out}\dagger}(x)\phi_{\pm'm}^{\text{out}}(x)\bigr]=\delta_{\pm\pm'}\delta_{nm}\,,\label{eq:orthonormal}
\end{equation}
and completeness
\begin{equation}
    \phi_{+n}^{\text{in}}(x)\phi_{+n}^{\text{in}\dagger}(y)+\phi_{-n}^{\text{in}}(x)\phi_{-n}^{\text{in}\dagger}(y)=\delta(\boldsymbol{x}-\boldsymbol{y})\,,\label{eq:complete}
\end{equation}
and likewise for the out state representation. The creation and annihilation operators act on their respective vacuum state, 
$a_{n}^{\textrm{out}},b_{m}^{\textrm{out}}|\textrm{out}\rangle=\langle\textrm{out}|a_{n}^{\textrm{out}\dagger},b_{m}^{\textrm{out}\dagger}=0$, 
and likewise for the in vacuum state.\footnote{We consider only vacuum states that act on spinor d.o.f., and do not consider dynamical photons, hence $|\textrm{out}\rangle \coloneqq |0;\textrm{out}\rangle$ and likewise for the in state.} The anti-commutation relations read $\{a_{n}^{\textrm{out}},a_{m}^{\textrm{out}\dagger}\}=\{b_{n}^{\textrm{out}},b_{m}^{\textrm{out}\dagger}\}=\delta_{nm}$, and likewise for the in state. Other anti-commutation relations are all zero, or in other words the other respective operators anti-commute. For further basic properties we refer the reader to Ref.~\cite{fradkin1991quantum}.

Let us next discuss the Green functions. We first distinguish their usage both in the determinant of matrix elements and as a mean expectation values, the former being in-out and the latter in-in. We write for the causal time ordered propagator and anti-time ordered propagator in the in-out representation as (here $z\coloneqq x-y$)
\begin{align}
    \label{eq:Sc_def}
    S^{c}(x,y) \coloneqq i\langle\mathcal{T}\psi(x)\bar{\psi}(y)\rrangle& \coloneqq \frac{i}{c_{v}}\langle\textrm{out}|\theta(z_{0})\psi(x)\bar{\psi}(y)-\theta(-z_{0})\bar{\psi}(y)\psi(x)|\textrm{in}\rangle\,,\\
    \label{eq:Scbar_def}
    S^{\bar{c}}(x,y) 
    \coloneqq i\llangle\bar{\mathcal{T}}\psi(x)\bar{\psi}(y)\rangle
    & \coloneqq \frac{i}{c_{v}^{*}}\langle\textrm{in}|\theta(-z_{0})\psi(x)\bar{\psi}(y)-\theta(z_{0})\bar{\psi}(y)\psi(x)|\textrm{out}\rangle\,,
\end{align}
with normalization, or equivalently the one-loop vacuum persistence.
given as 
\begin{equation}
    c_{v} \coloneqq\langle\textrm{out}|\textrm{in}\rangle\,.
\end{equation} 
We have made use of a compact notation for the in-out and out-in expectation value, i.e., $\langle\mathcal{O}\rrangle=c_{v}^{-1}\langle\textrm{out}|\mathcal{O}|\textrm{in}\rangle$
and $\llangle\mathcal{O}\rangle=c_{v}^{*-1}\langle\textrm{in}|\mathcal{O}|\textrm{out}\rangle$
for generic operator $\mathcal{O}$. We remark that by hermiticity
$[S^{c}(x,y)]{}^{\dagger}=-\gamma_{0}S^{\bar{c}}(y,x)\gamma_{0}$.
By contrast, one may likewise write the time-ordered and anti-time ordered in-in propagators as
\begin{align}
    S_{\text{in}}^{c}(x,y) & \coloneqq i\llangle\mathcal{T}\psi(x)\bar{\psi}(y)\rrangle\coloneqq i\langle\textrm{in}|\theta(z_{0})\psi(x)\bar{\psi}(y)-\theta(-z_{0})\bar{\psi}(y)\psi(x)|\textrm{in}\rangle\,,\\
    \Spropbar(x,y) &\coloneqq i\llangle\bar{\mathcal{T}}\psi(x)\bar{\psi}(y)\rrangle \coloneqq i\langle\textrm{in}|\theta(-z_{0})\psi(x)\bar{\psi}(y)-\theta(z_{0})\bar{\psi}(y)\psi(x)|\textrm{in}\rangle\,.
\end{align}
And we note here that by definition $\langle\text{in}|\text{in}\rangle=1$. The differential equations the propagator and anti-propagator solve
are respectively $(i\slashed{D}_{x}-m)\sprop(x,y)=-\delta(z)$ and
$(i\slashed{D}_{x}-m)\spropbar(x,y)=\delta(z)$, and likewise for $S_{\text{in}}^{c}(x,y)$ and $\Spropbar(x,y)$.

We will also need the in-out representation of the Wightman functions for both vacuum state orderings,
\begin{align}
    \label{eq:Wightman1}
    S^{>}(x,y) & =i\langle\psi(x)\bar{\psi}(y)\rrangle=\sprop(x,y)+\theta(-z_{0})G(x,y)\,,\\
    \label{eq:Wightman2}
    S^{<}(x,y) & =i\langle\bar{\psi}(y)\psi(x)\rrangle=-\sprop(x,y)+\theta(z_{0})G(x,y)\,,\\
    \label{eq:Wightman3}
    S^{\bar{>}}(x,y) & =i\llangle\psi(x)\bar{\psi}(y)\rangle=\spropbar(x,y)+\theta(z_{0})G(x,y)\,,\\
    \label{eq:Wightman4}
    S^{\bar{<}}(x,y) & =i\llangle\bar{\psi}(y)\psi(x)\rangle=-\spropbar(x,y)+\theta(-z_{0})G(x,y)\,,
\end{align}
where we have made use of the anti-commutation function, and also quantum mechanical Green function to express the Wightman functions in terms of the (anti-)causal Green functions. The anti-commutation function does not require any vacuum states and reads
\begin{equation}
    G(x,y)=i\{\psi(x),\bar{\psi}(y)\}\,;\label{eq:anticommutation}
\end{equation}
it satisfies the initial condition $G(x,y)|_{x_{0}=y_{0}}=i\gamma_{0}\delta(\boldsymbol{x}-\boldsymbol{y})$.
For completeness, similar formulations exist for the in-in construction:
\begin{align}
    S_{\textrm{in}}^{>}(x,y) & =i\llangle\psi(x)\bar{\psi}(y)\rrangle=\Sprop(x,y)+\theta(-z_{0})G(x,y)\,,\label{eq:S_in_>}\\
    S_{\textrm{in}}^{<}(x,y) & =i\llangle\bar{\psi}(y)\psi(x)\rrangle=-\Sprop(x,y)+\theta(z_{0})G(x,y)\,.\label{eq:S_in_<}
\end{align}
The Wightman and anti-commutation functions satisfy $(i\slashed{D}_{x}-m)S(x,y)=0$ for $S=S^{\lessgtr},S^{\bar{\lessgtr}}S_{\textrm{in}}^{\lessgtr},G$. $G$ has several important properties including serving as the time-translation operator of the fermion operator, i.e., with implicit sum over Dirac indices for common coordinates we have
\begin{equation}
    \int\! d^{3}y\,\psi^{\dagger}(y)G(y,x)\gamma_{0}=i\psi^{\dagger}(x)\,.
\end{equation}
We can, in spirit to those in a SK formalism, also write in-out statistical and spectral propagators respectively as
\begin{align}
    K(x,y) & =\frac{1}{2}[\sprop(x,y)+\spropbar(x,y)]\,,\label{eq:K_io}\\
    \rho(x,y) & =\frac{1}{2}[\sprop(x,y)-\spropbar(x,y)]\,.\label{eq:rho_io}
\end{align}
One also has with in-in vacuum expectation values statistical and spectral propagators
\begin{align}
    K_{\text{in}}(x,y) & =\frac{1}{2}[\Sprop(x,y)+\Spropbar(x,y)]\,,\\
    \rho_{\text{in}}(x,y) & =\frac{1}{2}[\Sprop(x,y)-\Spropbar(x,y)]
\end{align}
that do agree with, or are proportional to, the conventional propagators employed in the SK closed time path formalism. All in-in propagators whether of the causal, Wightman, or statistical or spectral variety agree with their in-out or out-in formulation when the vacuum persistence criteria is unity, i.e., $|\textrm{out}\rangle =|\textrm{in}\rangle$.

Our aim is to express more complicated, from a standard QFT treatment, in-in expressions in terms of first-quantized functions.
Here let us explore the known expressions, and some immediate extensions by virtue of the above formulae, in the first-quantized or worldline formalism that may serve as the basic building blocks. Let us begin with the well-known formulation of the time-ordered causal propagator, Eq.~\eqref{eq:Sc_def}, in an electromagnetic background~\cite{Schwinger:1951nm} written as formal inverse matrix element and with Schwinger propertime
\begin{equation}
    S^{c}(x,y)  =\langle x|\frac{-1}{i\hat{\slashed{D}}-m+i\epsilon}|y\rangle
     =(i\slashed{D}_{x}+m)\int_{0^{+}}^{\infty}\!ds\,e^{-\epsilon s}g(x,y,s)\,,\label{eq:Smatrix_element}
\end{equation}
where the Schwinger propertime kernel is given by
\begin{equation}
    g(x,y,s)\coloneqq i\langle x|e^{-i\hat{H}s}|y\rangle\,,
\end{equation}
with Hamiltonian for the quadratic propertime Dirac operator $\hat{H}\coloneqq\hat{\slashed{D}}{}^{2}+m^{2}$. The propertime kernel obeys $g(x,y,0^{+})=i\delta(x-y)$ and $g(x,y,0^{-})=-i\delta(x-y)$. The kernel of course has a convenient worldline representation:
\begin{equation}
    g(x,y,s)=i\int\mathcal{D}x\,e^{i\int_{0}^{s}d\tau[-m^2-\frac{\dot{x}^{2}}{4}-eA\cdot \dot{x}]}\mathcal{P}e^{-\frac{i}{2}\int_{0}^{s}d\tau\, eF\cdot\sigma}\,,
    \label{eq:worldline_pi}
\end{equation}
with boundary conditions as $x(0)=y$ and $x(s)=x$. Here, $\mathcal{P}$ denotes a path ordering in Schwinger propertime $s$. In this work we do not explicitly evaluate worldline path integral expressions apart from a few examples; however, it is tacitly assumed a worldline representation can always be had from the above worldline kernel. The anti-causal propagator, in Eq.~\eqref{eq:Scbar_def}, can simply be found from the Hermitian conjugate of Eq.~\eqref{eq:Smatrix_element}, or alternatively may be found from the SK formulation as we will illustrate below, and whose propertime lies in the negative direction,
\begin{align}
S^{\bar{c}}(x,y) =\langle x|\frac{1}{i\hat{\slashed{D}}-m-i\epsilon}|y\rangle
  =(i\slashed{D}_{x}+m)\int_{0^{-}}^{-\infty}\!ds\,e^{\epsilon s}g(x,y,s)\,.
  \label{eq:Sbarmatrix_element}
\end{align}
In what follows for brevity we will leave the $\epsilon$ factors guaranteeing convergence in the IR implicit in the integral symbol unless clarity is needed, i.e., $\int_{0^{+}}^{\infty}ds\,e^{-\epsilon s}\rightarrow\int_{0^{+}}^{\infty}ds$ and $\int_{0^{-}}^{-\infty}ds\,e^{\epsilon s}\rightarrow\int_{0^{-}}^{-\infty}ds$.

\begin{figure}[t!]
    \centering
    \begin{subfigure}[t]{0.4\textwidth}
        \centering
        \includegraphics[height=1.1in]{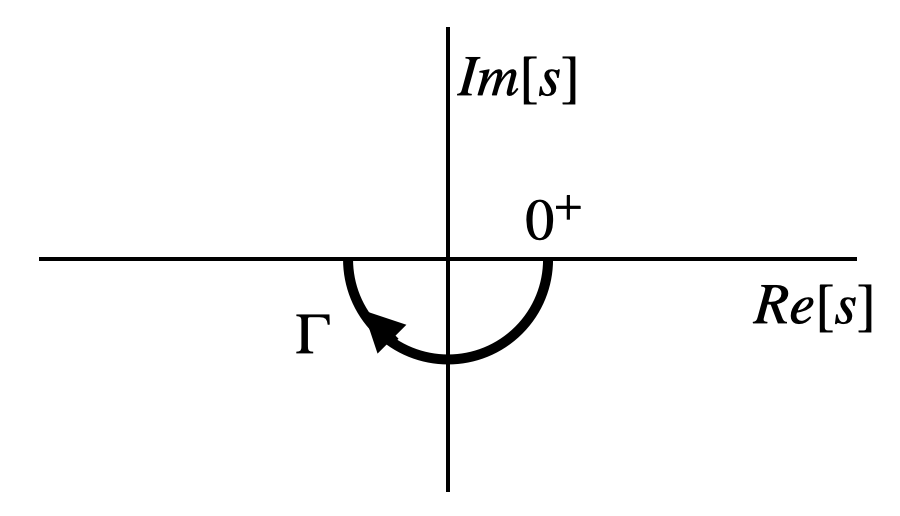}
        \caption{}
        \label{fig:Gamma}
    \end{subfigure}%
    ~
    \begin{subfigure}[t]{0.6\textwidth}
        \centering
        \includegraphics[height=1.45in]{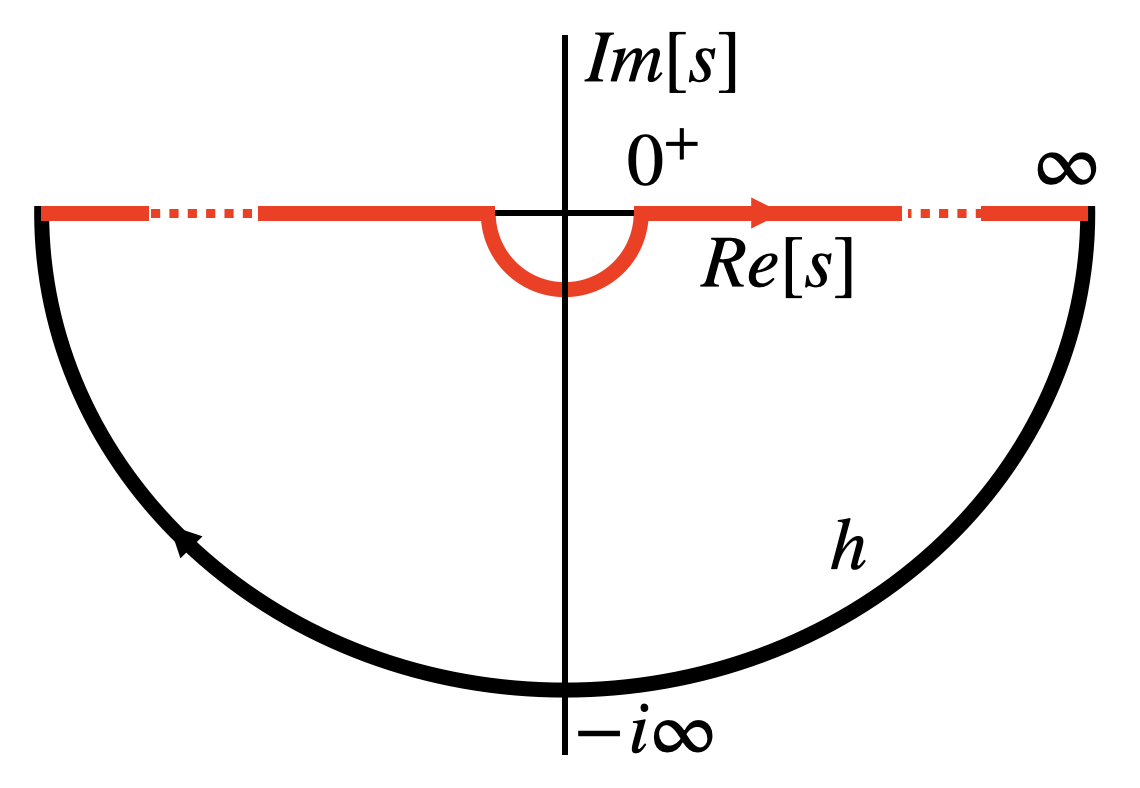}
        \caption{}
        \label{fig:h}
    \end{subfigure}
    \caption{(a) Contour for the Schwinger propertime kernel given for the anti-commutation function, $G(x,y)$, in Eq.~\eqref{eq:anticommutationWL}. (b) Contour for the density function serving to receive contributions from all singularities and critical points (poles, branch cuts, saddles) associated with Schwinger pair production in the imaginary Schwinger propertime plane. The contour of $\rho_h(x,y)$, in Eq.~\eqref{eq:rhoh}, is shown in red. The closed contour includes the contour in black, which is permissible for fields in the absence of branch cuts that might otherwise forbid the closure.}
\end{figure}

Next we look at the anti-commutation function, in Eq.~\eqref{eq:anticommutation}. Ref.~\cite{Gavrilov:1996ak} has demonstrated--see detailed derivation therein--that for arbitrary electromagnetic gauge field background the function is entirely expressible in Schwinger propertime as
\begin{equation}
    G(x,y)=(i\slashed{D}_{x}+m)\mathrm{sgn}(z_{0})\int_{\Gamma}\!ds\,g(x,y,s)\,,
    \label{eq:anticommutationWL}
\end{equation}
where $\Gamma$ denotes a clockwise half semicircle contour about
the origin in propertime that goes from $s\rightarrow 0^{+}$ to $s\rightarrow 0^{-}$ extending in the negative imaginary propertime plane; see Fig.~\ref{fig:Gamma}. It can be shown that under the $\Gamma$ contour $\int_\Gamma ds\,g(x,y,s)$ vanishes for coordinate arguments outside the light cone for $z^2<0$~\cite{fradkin1991quantum}. Thus the $\mathrm{sgn}(z_0)$ is commutable through the $\slashed{D}$ factor for arguments in which $\boldsymbol{z}^2>0$. Note that with the anti-commutation function no IR $\epsilon$ factor is required; however, since $s$ is about the origin an $e^{\pm\epsilon s}$ may be introduced under the integral to deform the contour as appropriate. With the propertime formulation of the anti-commutation function, in Eq.~\eqref{eq:anticommutation}, and the (anti-)causal propagators, it can be seen that all the Wightman functions, in Eqs.~\eqref{eq:Wightman1}-\eqref{eq:Wightman4}, and the statistical function in Eq.~\eqref{eq:K_io} and spectral function in Eq.~\eqref{eq:rho_io}, all have propertime formulations. Of particular importance is the spectral function.

Let us finally examine the spectral function in its propertime representation. One can see that both the causal and anti-causal propagators cover the entire real Schwinger propertime line with exception of the origin. To extend over the entire real line we make use of the half semicircle contour given in the anti-commutation function, $G(x,y)$, rewriting the spectral function as follows:
\begin{align}
    \rho(x,y) & =(i\slashed{D}_{x}+m)\frac{1}{2}\Bigl\{\int_{0^+}^{\infty}ds\,e^{-\epsilon s}\theta(s)+\int_{-\infty}^{0^-}ds\,e^{\epsilon s}\theta(-s)-\int_{\Gamma}+\int_{\Gamma}\Bigr\} g(x,y,s) \nonumber \\
    & =(i\slashed{D}_{x}+m)\frac{1}{2}\Bigl\{\int_{h}ds\,[\theta(s)e^{-\epsilon s}+\theta(-s)e^{\epsilon s}]+\int_{\Gamma}\Bigr\} g(x,y,s)\,,
\end{align}
where the contour, $h$, now covers the entire real axis of Schwinger propertime; see Fig.~\ref{fig:h}. In the absence of branch cuts that might otherwise forbid it, one may close the contour to encircle the entire imaginary complex plane. However, for many practical computations, it may prove advantageous to deform the contour so as to pass over saddle point, e.g., the worldline instanton method~\cite{Affleck:1981bma,Dunne:2005sx}. Again, let us absorb the IR convergence factor in the integral symbol itself, i.e., 
$\int_{h}ds\,[\theta(s)e^{-\epsilon s}+\theta(-s)e^{\epsilon s}]\rightarrow\int_{h}ds$. Then let us denote this portion of the spectral function as 
\begin{equation}
    \rho_{h}(x,y)=(i\slashed{D}_{x}+m)\frac{1}{2}\int_{h}\!ds\,g(x,y,s)\,,
    \label{eq:rhoh}
\end{equation}
or a vacuum instability Green function. Consider the imaginary part of the effective action written in propertime written suggestively with $h$, 
\begin{equation}
    2\mathrm{Im}\Gamma=2\mathrm{Im}(-i \ln c_v)=\Gamma-\Gamma^*=\mathrm{tr}\frac{1}{2}\int d^4x\int_h \frac{ds}{s} \,g(x,x,s)\,.
    \label{eq:ImGamma}
\end{equation}
Then, for example in a homogeneous electric field with strength $E$, we see that $h$ serves to pick up the poles in $g(x,x,s)$ at $s=-in/eE$, which are associated with the Schwinger effect, or a vacuum instability. For more general fields, the contour can receive contributions from all singularities and critical points associated with a vacuum instability. With this analogy we see that $\rho_h$ is the Green function equivalent of the imaginary part of the effective action; the difference amounts to a factor of $s^{-1}$ in the integrand. Expressing the spectral function as $\rho(x,y)=\rho_h(x,y)+2\mathrm{sgn}(z_0)G(x,y)$ we can see that the vacuum instability Green function expressed in terms of in out states is
\begin{equation}
    \rho_{h}(x,y) =\frac{i}{4}\Bigl\{\langle[\psi(x),\bar{\psi}(y)]\rrangle-\llangle[\psi(x),\bar{\psi}(y)]\rangle\Bigr\}\,,
\end{equation}
which we can see vanishes when $\langle\text{in}|=\langle\text{out}|$. We finally remark that the function satisfies the differential equation $(i\slashed{D}_{x}-m)\rho_{h}(x,y)=0$.

%%%%%%%%%%%%%%%%%%%%%%%%%%%%%%%%%%%%%%%%%%%%%%%%%%%%%%%%%%%%%%%%%%%
\section{In-in worldline formalism}
\label{sec:in-in}

Having explored some of the basic worldline building blocks, let us turn our attention to the formulation of in-in observables using them. Our goal in this section is to write down an in-in partition function and propagator, and hence related observables, entirely using the in-out propagators, $\sprop$~\eqref{eq:Smatrix_element}, $\spropbar$~\eqref{eq:Sbarmatrix_element}, the anti-commutation function, $G$~\eqref{eq:anticommutationWL}, and vacuum instability Green function, $\rho_h$~\eqref{eq:rhoh}, which are all well-defined in a first-quantized picture. We will accomplish this two equivalent ways, and in so doing draw connections between the two; the ways include a derivation from a Bogoliubov coefficient formalism~\cite{Gitman:1977ne,Fradkin:1981sc,fradkin1991quantum} and from the SK closed-time path formalism~\cite{Schwinger:1960qe,Keldysh:1964ud}. Let us begin with the former.

\subsection{Derivation from Bogoliubov coefficients}

To produce an in-in construction of the more well-known matrix element in-out worldline construction, let us first make use of an operator formalism. To begin, we note that the asymptotic creation and annihilation operators of are related to one another through a unitary transformation \cite{Gitman:1977ne,fradkin1991quantum}, here written without operators for later convenience,
\begin{align}
    U & =e^{-a_{n}^{\text{out}\dagger}\langle a_{n}^{\text{out}}b_{m}^{\text{out}}\rrangle b_{m}^{\text{out}\dagger}}e^{-a_{n}^{\text{out}}[\ln\langle a^{\text{out}}a^{\text{in}\dagger}\rrangle^{T}]_{nm}a_{m}^{\text{out}\dagger}}e^{b_{n}^{\text{out}\dagger}[\ln\langle b^{\text{out}}b^{in\dagger}\rrangle]_{nm}b_{m}^{\text{out}}}e^{-b_{n}^{\text{out}}\langle b_{n}^{in\dagger}a_{m}^{in\dagger}\rrangle a_{m}^{\text{out}}}\,,\\
    U^{\dagger} & =e^{-a_{n}^{\text{out}\dagger}\llangle a_{n}^{in}b_{m}^{in}\rangle b_{m}^{\text{out}\dagger}}e^{b_{n}^{\text{out}\dagger}[\ln\llangle b^{in}b^{\text{out}\dagger}\rangle]_{nm}b_{m}^{\text{out}}}e^{-a_{n}^{\text{out}}[\ln\llangle a^{\text{in}}a^{\text{out}\dagger}\rangle]_{nm}a_{m}^{\text{out}\dagger}}e^{b_{n}^{\text{out}}\llangle b_{n}^{\text{out}\dagger}a_{m}^{\text{out}\dagger}\rangle a_{m}^{\text{out}}}\,.
\end{align}
Relevant to our purposes, the unitary transform relates the in and out vacuum states as $\langle\textrm{in}|=\langle\text{out}|U^{\dagger}$.
In this way, one may define the out-in partition function in terms of creation and annihilation operators as
\begin{equation}
c_{v}^{*} =\langle\text{out}|e^{-a^{\text{out}}\ln\llangle a^{\text{in}}a^{\text{out}\dagger}\rangle a^{\text{out}\dagger}}|\textrm{out}\rangle =e^{\mathrm{tr}\ln\llangle a^{\text{in}}a^{\text{out\ensuremath{\dagger}}}\rangle} =\mathrm{det}\llangle a^{\text{in}}a^{\text{out\ensuremath{\dagger}}}\rangle\,,
\end{equation}
where we have left implicit the sum over eigenvalues. This then enables the in vacuum state to be written as~\cite{fradkin1991quantum}
\begin{align}
\langle\textrm{in}| & =\langle\text{out}|e^{-a^{\text{out}}\ln\llangle a^{\text{in}}a^{\text{out}}\rangle a^{\text{out}\dagger}}e^{a^{\text{out}}b^{\text{out}}\llangle b^{\text{out}\dagger}a^{\text{out}\dagger}\rangle}=c_{v}^{*}\langle\text{out}|e^{a^{\text{out}}b^{\text{out}}\llangle b^{\text{out}\dagger}a^{\text{out}\dagger}\rangle} \nonumber \\
 & =c_{v}^{*}\sum\limits _{N=0}^{\infty}\frac{1}{N!^{2}}\llangle a_{m_{N}}^{\text{out}\dagger}...\,a_{m_{1}}^{\text{out}\dagger}b_{n_{N}}^{\text{out}\dagger}...\,b_{n_{1}}^{\text{out}\dagger}\rangle\langle\text{out}|b_{n_{1}}^{\text{out}}...\,b_{n_{N}}^{\text{out}}a_{m_{1}}^{\text{out}}...\,a_{m_{N}}^{\text{out}}\,,
\end{align}
where in the last line, for later convenience, we have expanded the exponential, and shown the inverse operation of the Wick contractions.  Using the above we can see that in-in observables, e.g., $\llangle\mathcal{O}\rrangle$ for operator $\mathcal{O}$,
are entirely expressible in terms of in-out matrix elements, which in turn are then expressible in terms of worldline correlation functions. 

Let us begin with the simplest object, the vacuum non-persistence.
This actually follows from the definition of the in-in vacuum normalization upon insertion of a complete set of out states. Let us express this quantity in anticipation to a generating functional. It can be found that the only non-vanishing contributions are matrix elements where any number of particle antiparticle pairs are produced from the vacuum:
\begin{equation}
    \langle\textrm{in}|\textrm{in}\rangle=\sum\limits _{N=0}^{\infty} P_{N}=1\,,
\end{equation}
where the probability for $N$ pairs of particles to be produced in
the out state summed over all possible eigenstates is represented
by
\begin{equation}
    P_{N}\coloneqq\frac{|c_{v}|^{2}}{N!^{2}}|\langle a_{m_{N}}^{\textrm{out}}...\,a_{m_{1}}^{\textrm{out}}b_{n_{N}}^{\textrm{out}}...\,b_{n_{1}}^{\textrm{out}}\rrangle|^{2}\,.
\end{equation}
Again the various $m_{1},n_{1}...,m_{N},n_{N}$ are implicitly summed over with their respective hermitian conjugate. $P_0=|c_v|^2$. Rearranging the above by moving the $N=0$ contribution to the left we can see how the in-in generating functional contains information of the vacuum non-persistence,
\begin{equation}
    1-|c_{v}|^{2} =\llangle b_{n}^{\textrm{out}\dagger}a_{m}^{\textrm{out}\dagger}\rangle\langle a_{m}^{\textrm{out}}b_{n}^{\textrm{out}}\rrangle|c_{v}|^{2}+\frac{1}{2!^{2}}|\langle a_{m_{2}}^{\textrm{out}}a_{m_{1}}^{\textrm{out}}b_{n_{2}}^{\textrm{out}}b_{n_{1}}^{\textrm{out}}\rrangle|^{2}|c_{v}|^{2}+...\,.
 \label{eq:nonpersistance}
\end{equation}
The vacuum non-persistence is measuring the probability for any number of pairs of particles to be produced from the vacuum. Matrix elements here obey the usual Wick contractions; see Ref.~\cite{fradkin1991quantum} for details. 

To get a better sense of how the worldline construction fits into the above non-persistence written in terms of probabilities of pair produced in the out vacuum, let us first examine the simplest case in which only a single particle anti-particle pair appear in the vacuum out
state; this is given as
\begin{equation}
    P_{1} =|c_{v}|^{2}\langle b_{m}^{out\,\dagger}a_{n}^{out\,\dagger}\rrangle\llangle a_{n}^{out}b_{m}^{out}\rangle\,.
    \label{eq:P_1_def}
\end{equation}
Then using the definitions in Eqs.~\eqref{eq:inout_field_def}-\eqref{eq:complete}, notice that we may write the above with fermion operators (with Greek indices representing Dirac indices) as
\begin{equation}
    P_{1} =-|c_{v}|^{2}\int d^{3}xd^{3}y\langle\psi_{\alpha}^{\dagger}(y)\psi_{\beta}(x)\rrangle\llangle\psi_{\alpha}(y)\psi_{\beta}^{\dagger}(x)\rangle
    \label{eq:P_1_L}
\end{equation}
for any time $x_0$, $y_0$. Or, Eq.~\eqref{eq:P_1_def} may also likewise be written as 
\begin{equation}
    P_{1}=-|c_{v}|^{2}\int d^{3}xd^{3}y\langle\psi_{\beta}(x)\psi_{\alpha}^{\dagger}(y)\rrangle\llangle\psi_{\beta}^{\dagger}(x)\psi_{\alpha}(y)\rangle\,.
    \label{eq:P_1_R}
\end{equation}
Then it is convenient for later purposes to organize the time ordering of each so that
\begin{align}
    P_{1}  
    & =-|c_{v}|^{2}\mathrm{tr}\int d^{3}xd^{3}y\Bigl[\theta(-z_{0})\sprop(x,y)\gamma_{0}S^{\bar{>}}(y,x)\gamma_{0}-\theta(z_{0})\sprop(x,y)\gamma_{0}S^{\bar{<}}(y,x)\gamma_{0}\Bigr]
    \nonumber \\
    & =-|c_{v}|^{2}\int d^{3}xd^{3}y\,\mathrm{tr}\sprop(x,y)\Delta(y,x)\,,
\end{align}
where we have made use of the fact that $\theta(z_0)=\theta(z_0)^{2}$. We also have the combined propagator
\begin{equation}
    \label{eq:Delta_1}
    \Delta(x,y)\coloneqq\gamma_{0}[\spropbar(x,y)+\sgn(z_{0})G(x,y)]\gamma_{0}\,.
\end{equation}
The above construction is not unique, and other such equivalent ones may be written down, in particular we mention that the $\Delta$ function
may be represented in terms of $\rho_{h}$ as
\begin{equation}
    \Delta(x,y) =\gamma_{0}[\sprop(x,y)-2\rho_{h}(x,y)]\gamma_{0}\,,
    \label{eq:Delta}
\end{equation}
taking that $\mathrm{sgn}(z_0)^2=1$ and the fact that the $G(x,y)$ disappear in the $x_0\leftrightharpoons y_0$ limit. The $\Delta$ function we will go onto show isolates the vacuum instability condition, and will play the role of an effective interaction vertex in the in-out generating functional to describe an in-in setting.

Having examined the case of just one pair of particles being produced from the vacuum let us now examine the full case, predicting all possibilities of pair production. Of course we know the final result is simply the vacuum non-persistence given in Eq.~\eqref{eq:nonpersistance}. However, this exercise is important to not only validate our construction, but also will serve as a valuable introduction for calculations to come later. To begin let us write the in-in partition function as 
\begin{equation}
    \langle\textrm{in}|\textrm{in}\rangle=|c_{v}|^{2}\langle\exp(a_{n}^{\textrm{out}}b_{m}^{\textrm{out}}\llangle b_{m}^{\textrm{out}\dagger}a_{n}^{\textrm{out}\dagger}\rangle)\rrangle\,.
    \label{eq:Z_n_full}
\end{equation}
Again we proceed by rewriting the creation and annihilation operators
in their fermion operator representation. To do so let us examine
an $N$-th order term in an expansion of the exponential:
\begin{align}
    P_{N} & =\frac{|c_v|^2}{N!}\langle a_{n1}^{\textrm{out}}b_{m1}^{\textrm{out}}a_{n2}^{\textrm{out}}b_{m2}^{\textrm{out}}...a_{nN}^{\textrm{out}}b_{mN}^{\textrm{out}}\rrangle\llangle b_{m1}^{\textrm{out}\dagger}a_{n1}^{\textrm{out}\dagger}\rangle\llangle b_{m2}^{\textrm{out}\dagger}a_{n2}^{\textrm{out}\dagger}\rangle...\llangle b_{mN}^{\textrm{out}\dagger}a_{nN}^{\textrm{out}\dagger}\rangle\nonumber\\
    & =\frac{|c_v|^2}{N!}(-1)^{\frac{N(N+1)}{2}}\langle b_{m1}^{\textrm{out}}b_{m2}^{\textrm{out}}...b_{mN}^{\textrm{out}}a_{n1}^{\textrm{out}}a_{n2}^{\textrm{out}}...a_{nN}^{\textrm{out}}\rrangle\llangle b_{m1}^{\textrm{out}\dagger}a_{n1}^{\textrm{out}\dagger}\rangle\llangle b_{m2}^{\textrm{out}\dagger}a_{n2}^{\textrm{out}\dagger}\rangle...\llangle b_{mN}^{\textrm{out}\dagger}a_{nN}^{\textrm{out}\dagger}\rangle\nonumber\\
    & =\frac{|c_v|^2}{N!}(-1)^{\frac{N(N+1)}{2}}\int d^{3}x_{1,2,...,N}d^{3}y_{1,2,...,N}\langle\bar{\psi}(x_{1})\bar{\psi}(x_{2})...\bar{\psi}(x_{N})\psi(y_{1})\psi(y_{2})...\psi(y_{N})\rrangle\nonumber \nonumber\\
    & \quad\times\llangle\psi(x_{1})\bar{\psi}(y_{1})\rangle\llangle\psi(x_{2})\bar{\psi}(y_{2})\rangle...\llangle\psi(x_{N})\bar{\psi}(y_{N})\rangle\,,
    \label{eq:P_N_L}
\end{align}
where in the last line we have kept Dirac index contraction implicit
with it implied for common coordinates. One may equally well write
the equivalent formulation with operators interchanged:
\begin{align}
    P_{N} & =\frac{|c_v|^2}{N!}(-1)^{\frac{N(N+1)}{2}}\int d^{3}x_{1,2,...,N}d^{3}y_{1,2,...,N}\langle\psi(y_{1})\psi(y_{2})...\psi(y_{N})\bar{\psi}(x_{1})\bar{\psi}(x_{2})...\bar{\psi}(x_{N})\rrangle\nonumber \\
    & \quad\times\llangle\bar{\psi}(y_{1})\psi(x_{1})\rangle\llangle\bar{\psi}(y_{2})\psi(x_{2})\rangle...\llangle\bar{\psi}(y_{N})\psi(x_{N})\rangle\,.
    \label{eq:P_N_R}
\end{align}
Next to make the connection to the casual in-out propagators whose
worldline construction is well-established we use a common time for
all $(x_{N})_{0}$ and for all $(y_{N})_{0}$. We do this for both `halves' of $\theta(\pm z_0)$; this allows us to rewrite the correlators using a causal time-ordering since they are already time-ordered as they appear in Eqs.~\eqref{eq:P_N_L} and ~\eqref{eq:P_N_R} for $z_0>0$ and $z_0<0$ respectively.
\begin{align}
    P_{N} & 
    =\theta(z_{0})\times \textrm{Eq.~}\eqref{eq:P_N_L} +\theta(-z_{0})\times \textrm{Eq.~}\eqref{eq:P_N_R}\\
    & =\frac{1}{N!}\int d^{3}x_{1,2,...,N}d^{3}y_{1,2,...,N}\langle\mathcal{T}\psi(y_{1})\bar{\psi}(x_{1})\psi(y_{2})\bar{\psi}(x_{2})...\psi(y_{N})\bar{\psi}(x_{N})\rrangle\notag\\
    & \quad\times\prod_{n=1}^{N}\theta(z_{n0})\llangle\psi(x_{1})\bar{\psi}(y_{1})\rangle\llangle\psi(x_{2})\bar{\psi}(y_{2})\rangle...\llangle\psi(x_{N})\bar{\psi}(y_{N})\rangle\,.\notag\\
    & +\frac{(-1)^{N}}{N!}\int d^{3}x_{1,2,...,N}d^{3}y_{1,2,...,N}\langle\mathcal{T}\psi(y_{1})\bar{\psi}(x_{1})\psi(y_{2})\bar{\psi}(x_{2})...\psi(y_{N})\bar{\psi}(x_{N})\rrangle\notag\\
    & \quad\times\prod_{n=1}^{N}\theta(-z_{n0})\llangle\bar{\psi}(y_{1})\psi(x_{1})\rangle\llangle\bar{\psi}(y_{2})\psi(x_{2})\rangle...\llangle\bar{\psi}(y_{N})\psi(x_{N})\rangle
\end{align}
Then again remarking that $\theta(z_{0})\theta(-z_{0})=0$ we can
collect all the terms in the exponent to find
\begin{align}
    \langle\textrm{in}|\textrm{in}\rangle  & =|c_{v}|^{2}\langle  \mathcal{T}\exp\Bigl(i\int d^{3}xd^{3}y\,\bar{\psi}(x)\gamma_{0}[\theta(z_{0})S^{\bar{>}}(x,y)-\theta(-z_{0})S^{\bar{<}}(x,y)]\gamma_{0}\psi(y)\Bigr)\rrangle \nonumber \\
    & =|c_{v}|^{2}\langle\mathcal{T}\exp\Bigl(i\int d^{3}xd^{3}y\,\bar{\psi}(x)\Delta(x,y)\psi(y)\Bigr)\rrangle\,.
    \label{eq:in-in_gen}
\end{align}
We emphasize that the common times here, $x_{0}$ and $y_{0}$, are
valid for any real value. 

Notice that in the above derivation of the in-in generating functional, we never used the enclosing in state, for example as it appears in Eq.~\eqref{eq:in-in_gen}. Therefore all of the above arguments may equally well be applied to a formulation of the in state in terms of squeezed states represented with $\Delta$ as
\begin{equation}
    \langle \textrm{in}|=c_{v}^*\langle\textrm{out}|\mathcal{T}\exp\Bigl(i\int d^{3}xd^{3}y\,\bar{\psi}(x)\Delta(x,y)\psi(y)\Bigr)\,.
\end{equation}
Therefore what we can gather from the above is that a similar structure exists for in-in bilinears, and other spinor operator insertions in between the in-in states. However, in order to maintain the time ordering the collective $x_0$ and $y_0$ must both be greater than the times of any inserted operators. A simple requirement satisfying any operator insertions is simply to take $x_0,\,y_0\to\infty$, and hence realizing the SK contour connection as asymptotic infinity. With such a choice spinor operators may be inserted into the time-ordering, and a representation for the in-in causal propagator in terms of in-out matrix elements may be found as
\begin{equation}
    S_{\text{in}}^{c}(x',y')=|c_{v}|^{2}\langle\mathcal{T}\exp\Bigl(\lim_{x_{0},y_{0}\rightarrow\infty}\!i\int d^{3}xd^{3}y\,\bar{\psi}(x)\Delta(x,y)\psi(y)\Bigr)\psi(x')\bar{\psi}(y')\rrangle\,.
    \label{eq:S_in^c_final}
\end{equation}
Taking it a step further, and written suggestively for later comparison purposes, let us write down an in-in generating functional as
\begin{equation}
    \mathcal{Z}_{\eta}^{++}=|c_{v}|^{2}\langle\mathcal{T}\exp\Bigl(\Bigl\{\lim_{x_{0},y_{0}\rightarrow\infty}\!i\int d^{3}xd^{3}y\,\bar{\psi}(x)\Delta(x,y)\psi(y)\Bigr\}+i\int d^{4}x[\bar{{\eta}}\psi+\bar{\psi}\eta]\Bigr)\rrangle\,.\label{eq:Z^++}
\end{equation}
The ``$++$" in SK parlance denotes that both $\eta$ and $\bar{\eta}$ lie on the top time contour and represent time-ordered operator insertions, as opposed to the mixed or ``$--$" representations. Naturally one has that $\mathcal{Z}_{\eta=0}^{++}=\langle\textrm{in}|\textrm{in}\rangle=1$.

%%%%%%%%%%%%%%%%%%%%%%%%%%%%%%%%%%%%%%%%%%%%%%%%%%%%%%%%%%%%%%%%%%%%%%%%
\subsection{Derivation from the Schwinger-Keldysh CTP formalism}
\label{sec:SK}

\begin{figure}
\centering
\includegraphics[scale=0.3]{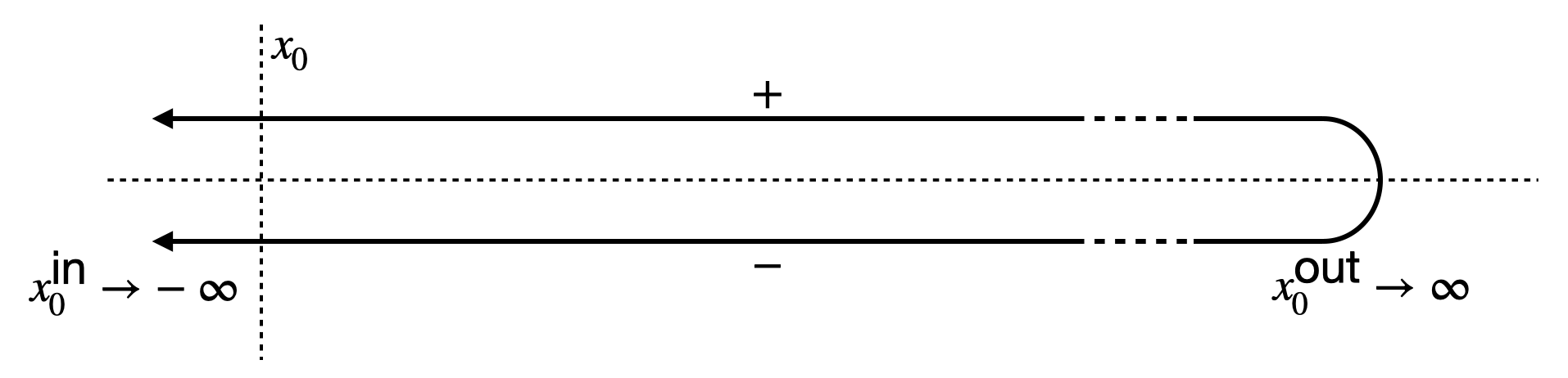}
\caption{
SK closed-time contour $\mathcal{C}$. Causal half, $+$, extends from $x_0^\text{in}\to-\infty$ to $x_0^\text{out}\to\infty$, and anti-causal half, $-$, in the opposite direction. Both $\pm$ paths are non-trivially connected at $x_0^\text{out}$, leading to for example for Dirac operators the following BC:
}
\label{fig:sk_contour}
\end{figure}

It is understood that the in-in formalism is equivalent to the SK formalism~\cite{PhysRevD.33.444}, and it is prudent that we establish this connection both to verify our finding above and to provide insight into the in-in construction. To begin let write down the well-known SK generating functional that includes all ``$\pm\pm$" components as
\begin{equation}    \mathcal{Z}_{\eta}=\frac{1}{\mathcal{Z}_{0}}\int\mathcal{D}\psi\mathcal{D}\bar{\psi}\,\exp\Bigl(i\int d^{3}x\int_{\mathcal{C}}dx^{0}\mathcal{L}[\psi,\overline{\psi},\eta,\overline{\eta}]\Bigr)\,,\label{eq:sk_functional-1}
\end{equation}
where the Dirac Lagrangian contains grassmann sources, and reads
$\mathcal{L}=\bar{\psi}(i\slashed{D}-m)\psi+\bar{{\eta}}\psi+\bar{\psi}\eta$.
The closed time path contour $\mathcal{C}$ in the SK formalism consists of two time contours that are both forward and reversed and connected at $x_{0}\rightarrow\infty$; see Fig.~\ref{fig:sk_contour}.
All the non-triviality of the contour lies in the boundary condition, i.e, $\psi_{+}(x_{0}\rightarrow\infty)=\psi_{-}(x_{0}\rightarrow\infty)$ for $\psi_{+}$ and $\psi_{-}$ lying on the top and bottom contours respectively, at temporal infinity, and this is because the connection there represents a sum over all out states~\cite{PhysRevD.33.444}. To further investigate the structure let us unpack the expression. To do so, let us represent the non-trivial boundary condition through a Lagrange multiplier or delta function connecting both branches of the contour at asymptotic infinity as
\begin{equation}
    \mathcal{Z}_{\eta}=\frac{1}{\mathcal{Z}_{0}}\int\mathcal{D}\psi_{\pm}\mathcal{D}\bar{\psi}_{\pm}\,\Bigl\{\lim_{x_{0}\rightarrow\infty}\delta[\psi_{+}(x)-\psi_{-}(x)]\delta[\bar{\psi}_{+}(x)-\bar{\psi}_{-}(x)]\Bigr\}\,e^{i\int d^{4}x\,[\mathcal{L}_+-\mathcal{L}_-]}\,.
\end{equation}
Here $\mathcal{L}_\pm$ represents the Dirac Lagrangian with $\psi\to\psi_\pm$ on either the $+$ or $-$ contour, and with $\eta\to\eta_\pm$. To represent the functional delta function in path integral form, we use an anti-commuting fermionic variables, $\lambda$ and $\bar{\lambda}$, obeying similar statistics as $\psi$. One may then write 
\begin{align}
    \mathcal{Z}_{\eta} & = \frac{1}{\mathcal{Z}_0}\int\mathcal{D}\psi_{\pm}\mathcal{D}\bar{\psi}_{\pm}\mathcal{D}\lambda \mathcal{D}\bar{\lambda}\,\exp\Bigl\{\lim_{x_{0}\rightarrow\infty}i\int d^{3}x[\bar{\lambda}(\psi_{+}-\psi_{-})+(\bar{\psi}_{+}-\bar{\psi}_{-})\lambda\Bigr\}
    e^{ i\int d^{4}x[\mathcal{L}_+-\mathcal{L}_-] }.
\end{align} 
With the above formulation in place in order to determine an effective action, and hence first quantized representation from the effective action, we must integrate out the various path integrals.

To make comparison to the expression found previously let us turn our attention to the case of the generating functional for causal propagators. Then we need only treat the case of the $+$ contour lying on top. We can do this by setting $\eta_{-}=\bar{\eta}_{-}=0$, then $\mathcal{Z}_{\eta_-=0}\to\mathcal{Z}_\eta^{++}$ with
\begin{align}
    \mathcal{Z}_{\eta}^{++} & =\frac{1}{\mathcal{Z}_0}\int\mathcal{D}\psi_{\pm}\mathcal{D}\bar{\psi}_{\pm} \mathcal{D}\lambda \mathcal{D}\bar{\lambda}\,\exp\Bigl\{\lim_{x_{0}\rightarrow\infty}i\int d^{3}x[\bar{\lambda}(\psi_{+}-\psi_{-})+(\bar{\psi}_{+}-\bar{\psi}_{-})\lambda\Bigr\}\nonumber \\
    & \quad\times \exp\Bigl\{i\int d^{4}x[\mathcal{L}_+-\bar{\psi}_{-}(i\slashed{D}-m)\psi_{-}]\Bigr\}\,.
\end{align}
Next, we evaluate the $\psi_{-}$ integrals, which can be accomplished with the following redefinitions: 
\begin{align}
    \psi_{-}'(x)&=\psi_{-}(x)+\lim_{y_{0}\rightarrow\infty}\int d^{3}y\spropbar(x,y)\lambda(y)\,,\\
    \bar{\psi}_{-}'(x)&=\bar{\psi}_{-}(x)+\lim_{y_{0}\rightarrow\infty}\int d^{3}y\bar{\lambda}(y)\spropbar(y,x)\,.
\end{align}
After inserting the above into the argument of the exponential, we can `complete the square;' we will pick up a determinant factor from the ensuing quadratic term on the $-$ contour that will yield a factor of $c_v^*$, and a non-local term from convolution of the additional $S\lambda$ factors in the redefinition. Completing the square we can then find
\begin{align}
    \mathcal{Z}_{\eta}^{++} & =\frac{c_{v}^{*}}{\mathcal{Z}_0}\int\mathcal{D}\psi\mathcal{D}\bar{\psi} \mathcal{D}\lambda \mathcal{D}\bar{\lambda}\,\exp\Bigl\{\lim_{x_{0}\rightarrow\infty}i\int d^{3}x[\bar{\lambda}\psi+\bar{\psi}\lambda]\Bigr\}\nonumber \\
    & \quad\times e^{i\int d^{4}x\bigl\{\bar{\psi}(i\slashed{D}-m)\psi+\bar{{\eta}}\psi+\bar{\psi}\eta\bigr\}+i\lim_{x_{0},y_{0}\rightarrow\infty}\int d^{3}x\int d^{3}y\bar{\lambda}(y)\spropbar(y,x)\lambda(x)}\,,
\end{align}
where we have went ahead and dropped the $+$ from the $\psi$ and $\eta$ variables. Notice now that we have a non-local term in the $\lambda$ variables, and it is treated in a coincident limit way. To integrate out $\lambda$ we must treat the asymptotic time dependence with care. We relax $x_0$ and $y_0$ allowing a coincident prescription, i.e., $\mathcal{Z}_\eta^{++}=\lim_{x_0,y_0\to\infty}\mathcal{Z}_{\eta}^{++}(x_0,y_0)$ with $x_0$ and $y_0$ now taking on finite values, where now we may judiciously treat either coincident limit as
\begin{align}
    \mathcal{Z}_{\eta}^{++}(x_0,y_0)&=\theta(z_{0})\frac{c_{v}^{*}}{\mathcal{Z}_0}\int\mathcal{D}\psi\mathcal{D}\bar{\psi}\mathcal{D}\lambda\mathcal{D}\bar{\lambda}\,\exp\Bigl\{ i\int d^{3}x[\bar{\lambda}_{y_{0}}\psi_{y_{0}}+\bar{\psi}_{x_{0}}\lambda_{x_{0}}]\Bigr\}\notag\\
	&\quad\times e^{i\int d^{4}x\Bigl\{\bar{\psi}(i\slashed{D}-m)\psi+\bar{{\eta}}\psi+\bar{\psi}\eta\Bigr\}-\int d^{3}x\int d^{3}y\bar{\lambda}_{y_{0}}(y)\llangle\psi_{y_{0}}(y)\bar{\psi}_{x_{0}}(x)\rangle\lambda_{x_{0}}(x)}\notag\\
	&+\theta(-z_{0})\frac{c_{v}^{*}}{\mathcal{Z}_0}\int\mathcal{D}\psi\mathcal{D}\bar{\psi}\mathcal{D}\lambda\mathcal{D}\bar{\lambda}\,\exp\Bigl\{ i\int d^{3}x[\bar{\lambda}_{y_0}\psi_{y_0}+\bar{\psi}_{x_0}\lambda_{x_0}]\Bigr\}\notag\\
	&\quad\times e^{i\int d^{4}x\bigl\{\bar{\psi}(i\slashed{D}-m)\psi+\bar{{\eta}}\psi+\bar{\psi}\eta\bigr\}+\int d^{3}x\int d^{3}y \bar{\lambda}_{y_0}(y)\llangle\bar{\psi}_{x_0}(x)\psi_{y_0}(y)\rangle\lambda_{x_0}(x)}\,,
    \label{eq:Z++2}
\end{align}
and we have made time dependence explicit with a subscript notation. Let us look at the top set of integrals with $\theta(z_0)$. There we perform the following shifts:
\begin{align}
    \lambda_{x_{0}}(x)&=\lambda'_{x_{0}}(x)+\int d^{3}x'\Delta_{x_{0}y_{0}}(x,x')\psi_{y_{0}}(x')\nonumber\\&=\lambda'_{x_{0}}(x)+i\int d^{3}x'\gamma_{0}\llangle\psi_{x_{0}}(x)\bar{\psi}_{y_{0}}(x')\rangle\gamma_{0}\psi_{y_{0}}(x')\,,\\\bar{\lambda}_{y_{0}}(y)&=\bar{\lambda}'_{y_{0}}(y)+\int d^{3}x''\bar{\psi}_{x_{0}}(x'')\Delta_{x_{0}y_{0}}(x'',y)\nonumber\\&=\bar{\lambda}'_{y_{0}}(y)+i\int d^{3}x''\psi_{x_{0}}^{\dagger}(x'')\llangle\psi_{x_{0}}(x'')\psi_{y_{0}}^{\dagger}(y)\rangle\,.
\end{align}
Then after arranging a few terms and making use of the identities in Eqs.~\eqref{eq:orthonormal} and~\eqref{eq:complete}, the affected parts of the action become
\begin{align}
    &i\int d^{3}x[\bar{\lambda}_{y_{0}}\psi_{y_{0}}+\bar{\psi}_{x_{0}}\lambda_{x_{0}}]-\int d^{3}x\int d^{3}y\bar{\lambda}_{y_{0}}(y)\llangle\psi_{y_{0}}(y)\bar{\psi}_{x_{0}}(x)\rangle\lambda_{x_{0}}(x)\nonumber\\
    \to& 
    -\int d^{3}x\int d^{3}y\psi_{x_{0}}^{\dagger}\llangle\psi_{x_{0}}(x)\psi_{y_{0}}^{\dagger}(y)\rangle\psi_{y_{0}}(y)-\int d^{3}x\int d^{3}y\bar{\lambda}'_{y_{0}}(y)\llangle\psi_{y_{0}}(y)\bar{\psi}_{x_{0}}(x)\rangle\lambda'_{x_{0}}(x)\notag\\
    &+i\int d^{3}y\int d^{3}x\bar{\lambda}'_{y_{0}}(y)\llangle\psi_{y_{0}}^{\dagger}(x)\psi_{y0}(y)\rangle\psi_{y_{0}}(x)+i\int d^{3}x\int d^{3}y\psi_{x_{0}}^{\dagger}(y)\llangle\bar{\psi}_{x_{0}}(x)\psi_{x_{0}}(y)\rangle\lambda'_{x_{0}}(x)\,.\notag
\end{align}
Notice, however, that the final two terms on the last line of the above are written in the order $\psi^\dagger \psi$ whereas the quadratic term in $\lambda$ has the order $\psi \psi^\dagger$. These terms are not time-ordered. The key point is that the only contractions in $\lambda$ and $\bar{\lambda}$ in the final two terms on the last line involve a convolution in which they will vanish, e.g., $\int d^{3}x\langle\psi^{\dagger}(x)\psi(y)\rrangle\langle\psi(x)\psi^{\dagger}(y')\rrangle=0$; therefore the two terms vanish. An analogous calculation for the $\theta(-z_0)$ terms in Eq.~\eqref{eq:Z++2}, with similar shifts, can also be accomplished. 

Gathering both sides, $\theta(\pm z_0)$, of~\eqref{eq:Z++2} we find after combining both into the exponent that
\begin{align}
    \mathcal{Z}_{\eta}^{++}(x_0,y_0)=&\frac{c_{v}^{*}}{\mathcal{Z}_0}\int\mathcal{D}\psi\mathcal{D}\bar{\psi}\mathcal{D}\lambda\mathcal{D}\bar{\lambda}\,\exp\Bigl\{ i\int d^{3}xd^3y\,\bar{\psi}(y)\Delta(y,x)\psi(x)\Bigr\}\notag\\
	&\times e^{i\int d^{4}x\bigl\{\bar{\psi}(i\slashed{D}-m)\psi+\bar{{\eta}}\psi+\bar{\psi}\eta\bigr\}-\int d^{3}x\int d^{3}y\bar{\lambda}(y)\spropbar(y,x)\lambda(x)}\,.
\end{align}
Then fixing the normalization to correctly reproduce the normalization of the in state as $\mathcal{Z}_0=\int d\lambda d\bar{\lambda}\exp[i\int d^{3}x\int d^{3}y\,\bar{\lambda}(y)\spropbar(y,x)\lambda(x)]$, and finally making explicit the coincident limit we find in exact agreement to Eq.~\eqref{eq:Z^++} in the path integral representation that
\begin{align}
    \mathcal{Z}_{\eta}^{++} & =c_{v}^{*}\int\mathcal{D}\psi\mathcal{D}\bar{\psi}\, e^{i\int d^{4}x\bigl\{\bar{\psi}(i\slashed{D}-m)\psi+\bar{{\eta}}\psi+\bar{\psi}\eta\bigr\}+ i\lim_{x_{0},y_{0}\to\infty}\int d^{3}xd^{3}y\,\bar{\psi}(y)\Delta(y,x)\psi(x)}\,.
\end{align}

Finally, while the causal $++$ generating functional is most convenient, let us remark on the other parts of the SK generating functional. A more general in-in generating functional includes paths on the $-+$, $+-$, and $--$ contours, making up a $2\times2$ matrix (one may equivalently use a retarded/advanced prescription). In which case one would find for the $2\times2$  propagator the causal propagator given in Eq.~\eqref{eq:S_in^c_final}. The $-+$ and $+-$ propagators, in Eqs.~\eqref{eq:S_in_>} and~\eqref{eq:S_in_<}, follow immediately using the anti-commutation function in Eq.~\eqref{eq:anticommutationWL}. Finally the in-in anti-causal propagator can be found from  $[S^{c}_\text{in}(x,y)]{}^{\dagger}=-\gamma_{0}S^{\bar{c}}_\text{in}(y,x)\gamma_{0}$, completing the components of $2\times2$  SK matrix propagator.

%%%%%%%%%%%%%%%%%%%%%%%%%%%%%%%%%%%%%%%%%%%%%%%%%%%%%%%%%%%%%%%
\section{Resummation and worldline total pair number probability}
\label{sec:pairs}

We are now in a position to evaluate the in-in partition function. In the absence of sources we will find that the vacuum persistence takes on a nice resummed form in which the probability for $N$ particle anti-particle pairs being created from the vacuum are expressible in first-quantized language. The resummation structure was first identified in Ref.~\cite{Copinger:2024pai}, in which the pair number probability was expressed in terms of amplitude using a coherent states formalism, and we generalize here to the case arbitrary fields, and emphasize the first-quantized representation of the pair number probability. Casting the pair number probability in terms of first-quantized expressions is key as the formalism benefits from an all-orders in the background field coupling prescription--sometimes necessary to extract non-perturbative information of the Schwinger effect in complicated background fields.

Let us evaluate the in-in partition function, 
\begin{align}
    \langle\textrm{in}|\textrm{in}\rangle=|c_{v}|^{2}\langle\mathcal{T}\exp\Bigl(i\int d^{3}xd^{3}y\bar{\psi}(x)\Delta(x,y)\psi(y)\Bigr)\rrangle\,,
\end{align}
where we emphasize that for the \textit{partition function} only we may allow $x_0$ and $y_0$ to take on any value. The argument of the exponent is quadratic and hence one may evaluate this exactly for all orders, however to demonstrate this let us perform the Wick contraction directly at each order for good measure. Here we adopt the simplified notation of  $\bar{\psi}\Delta\psi=\bar{\psi}_{x}\Delta_{xy}\psi_{y}\coloneqq\int d^{3}xd^{3}y\bar{\psi}(y)\Delta(y,x)\psi(x)$ with an implied integration over space and contraction of Dirac indices. Where no contraction is present we simply write $\psi_{x'}\coloneqq\psi(x')$. Then for an expansion of the exponential let us look at an $n$-th element:
\begin{align}
    \langle\mathcal{T}&\psi_{x1}\bar{\psi}_{y_{1}}...\psi_{x_{n}}\bar{\psi}_{y_{n}}\psi_{x'}\bar{\psi}_{y'}\rrangle 
    =\langle\mathcal{T}\boxed{\psi_{x1}\bar{\psi}_{y_{1}}...\psi_{x_{n}}\bar{\psi}_{y_{n}}}\,\contraction{}{\psi_{x'}}{}{\bar{\psi}_{y'}}\psi_{x'}\bar{\psi}_{y'}\rrangle\notag\\
    & +n\,\langle\mathcal{T}\boxed{\psi_{x1}\bar{\psi}_{y_{1}}...\psi_{x_{n-1}}\bar{\psi}_{y_{n-1}}}\,\contraction{\psi_{x_{n}}}{\bar{\psi}_{y_{n}}}{}{\psi_{x'}}\contraction[2ex]{}{\psi_{x_{n}}}{\bar{\psi}_{y_{n}}\psi_{x'}}{\bar{\psi}_{y'}}\psi_{x_{n}}\bar{\psi}_{y_{n}}\psi_{x'}\bar{\psi}_{y'}\rrangle\notag\\
    & +n(n-1)\,\langle\mathcal{T}\boxed{\psi_{x1}\bar{\psi}_{y_{1}}...\psi_{x_{n-2}}\bar{\psi}_{y_{n-2}}}\contraction{\psi_{x_{n}}}{\bar{\psi}_{y_{n-1}}}{}{\psi_{x_{n}}}\contraction{\psi_{x_{n-1}}\bar{\psi}_{y_{n-1}}\psi_{x_{n}}}{\bar{\psi}_{y_{n}}}{}{\psi_{x'}}\contraction[2ex]{}{\psi_{x_{n-1}}}{\bar{\psi}_{y_{n-1}}\psi_{x_{n}}\bar{\psi}_{y_{n}}\psi_{x'}}{\bar{\psi}_{y'}}\psi_{x_{n-1}}\bar{\psi}_{y_{n-1}}\psi_{x_{n}}\bar{\psi}_{y_{n}}\psi_{x'}\bar{\psi}_{y'}\rrangle+...\,.
\end{align}
Here the boxed operators represent all possible permutations of contractions. In the first line, we have contracted the unique $\psi_{x'}$ and $\psi_{y'}$. In the next line, contractions with $\psi_{x'}$ and $\psi_{y'}$ and any one of the $n$ fermions in the interaction term are accounted for. A similar structure exists when borrowing two or more sets of fermions from the interaction term. We then find for the in-in partition function the following:
\begin{align}
\langle\text{in}|\text{in}\rangle & =|c_{v}|^{2}\sum_{n=0}^{\infty}\frac{1}{n!}\langle\mathcal{T}[-i\psi_{x}\bar{\psi}_{y}\Delta_{yx}]^{n}\rrangle \nonumber\\
 & =|c_{v}|^{2}-|c_{v}|^{2}\sum_{n=1}^{\infty}\frac{1}{n!}\langle\mathcal{T}[-i\psi_{x}\bar{\psi}_{y}\Delta_{yx}]^{n-1}\rrangle\sprop_{xy}\Delta_{yx}\notag\\
 & \quad+|c_{v}|^{2}\sum_{n=2}^{\infty}\frac{1}{n!}(n-1)\langle\mathcal{T}[-i\psi_{x}\bar{\psi}_{y}\Delta_{yx}]^{n-2}\rrangle\sprop_{x'y'}\Delta_{y'x''}\sprop_{x''y''}\Delta_{y''x'}\,...\nonumber\\
 & =|c_{v}|^{2}-|c_{v}|^{2}\sum_{n=0}^{\infty}\frac{1}{(n+1)n!}\langle\mathcal{T}[-i\psi_{x}\bar{\psi}_{y}\Delta_{yx}]^{n}\rrangle\sprop_{xy}\Delta_{yx}\notag\\
 & \quad+|c_{v}|^{2}\sum_{n=0}^{\infty}\frac{1}{(n+2)n!}\langle\mathcal{T}[-i\psi_{x}\bar{\psi}_{y}\Delta_{yx}]^{n}\rrangle\sprop_{x'y'}\Delta_{y'x''}\sprop_{x''y''}\Delta_{y''x'}\,...\,.
\end{align}
We can see a pattern forming, and this is one of the Bell polynomials (see Ref.~\cite{Copinger:2024pai} for their appearance from a Furry expansion). We can resum the series into a compact form using a functional trace, $\mathrm{Tr}_3$, that acts on both the 3-dimensional spacial variable as well as the Dirac indices ($\mathrm{tr}$ as before); likewise common times are still assumed, i.e., $x^0_i=x^0$ and $y^0_i=y^0$ for all $i$. For example, we have for the following product chain
\begin{equation}
    \mathrm{Tr}_3[\sprop\Delta]^{n}\coloneqq\prod_{i=1}^{n}\int d^{3}x_{i}d^{3}y_{i}\mathrm{tr}[\sprop_{x_{1}y_{1}}\Delta_{y_{1}x_{2}}\sprop_{x_{2}y_{2}}\Delta_{y_{2}x_{3}}\,...\,\sprop_{x_{n}y_{n}}\Delta_{y_{n}x_{1}}]\Big|_{x_{i}^{0}=x_{j}^{0},\,y_{i}^{0}=y_{j}^{0}}\,.
\end{equation}
Then the in-in partition function can be represented as
\begin{align}
    \langle\text{in}|\text{in}\rangle 
    & =|c_{v}|^{2}\Bigl\{I-\mathrm{Tr}_3\sprop\Delta+\frac{1}{2!}\Bigl([\mathrm{Tr}_3\sprop\Delta]^{2}-\mathrm{Tr}_3[\sprop\Delta\sprop\Delta]\Bigr)+...\Bigr\}\\
    & =|c_{v}|^{2}\exp\Bigl[-\sum_{n=1}^{\infty}\frac{1}{n}\mathrm{Tr}_3[\sprop\Delta]^{n}\Bigr]\,,\label{eq:in-in_exp_Tr}
\end{align}
which is immediately identifiable in determinant form as
\begin{equation}
    \langle\text{in}|\text{in}\rangle =|c_{v}|^{2}\exp\mathrm{Tr}_3\ln[I-\sprop\Delta] =|c_{v}|^{2}\,\mathrm{Det}_3[I-\sprop\Delta]\,,
    \label{eq:det_3}
\end{equation}
from which we see the Bell polynomial form follows. Let us go ahead and express the total pair production probability using $\sum\limits _{N=0}^{\infty}P_{N}=\langle\text{in}|\text{in}\rangle$ as
\begin{equation}
    P_{N}=\frac{|c_{v}|^{2}}{N!} B_N\Bigl(  \mathrm{Tr}_3[\sprop\Delta]\,,\ldots\,,(N-1)!\mathrm{Tr}_3[\sprop\Delta]^N \Bigr)\,,
\end{equation}
from which one may cast into Schwinger propertime/worldline representation with ease. Here, $B_N$ is the $N$-th complete exponential Bell polynomial. There are different possible representations; we illustrate first the one with Eq.~\eqref{eq:Delta_1}, to emphasize a forward and backward pass in propertime:
\begin{equation}
    \sprop(x,y)\Delta(y,x')=(i\slashed{D}_{x}+m)\int_{0^{+}}^{\infty}ds\left[\int_{0^{-}}^{-\infty}+\int_\Gamma \right]ds'\,g(x,y,s)\gamma_{0}(i\slashed{D}_{x'}+m)g(y,x',s')\gamma_{0}\,,
    \label{eq:SDelta_1}
\end{equation}
which apart from the $\Gamma$ contour represents the hermitian conjugate of the matrix element causal propagator. One may also, using Eq.~\eqref{eq:Delta}, show the above using the closed contour over the entire imaginary plane:
\begin{equation}
    \sprop(x,y)\Delta(y,x')=-(i\slashed{D}_{x}+m)\int_{0^{+}}^{\infty}ds\int_h ds'\,g(x,y,s)\gamma_{0}(i\slashed{D}_{x'}+m)g(y,x',s')\gamma_{0}\,,
    \label{eq:SDelta_2}
\end{equation}
where we have applied the common time $x_0=x_0'$, and the identity,
\begin{equation}
    \int d^3y\,\sprop(x,y)\gamma_0\sprop(y,x')\gamma_0=[i\theta(x_{0}-y_{0})\theta(y_{0}-x'_{0})-i\theta(x'_{0}-y_{0})\theta(y_{0}-x_{0})]\sprop(x,x')\gamma^{0}\,,
    \label{eq:SS_cancel}
\end{equation}
found from the causal propagators' defining identities given in Eq.~\eqref{eq:orthonormal} and~\eqref{eq:complete}; the above vanishes for $x_0=x_0'$. The above forms are complete, however, for the case of the pair production probability we can make use of the determinant structure to further simplify the formulae. 

In order to simplify the determinant and hence pair production probability, let us make use of yet another form of the convolution of $\sprop\Delta$. First let us write down another identity:
\begin{equation}
    \mathrm{sgn}(y_0-x_0')\int d^3y\,\sprop(x,y)\gamma_0G(y,x')\gamma_0\Big|_{x'_0=y_0}\!=
    \theta(z_0)\langle\psi(x)\psi^{\dagger}(x')\rrangle+\theta(-z_0)\langle\psi^{\dagger}(x')\psi(x)\rrangle\,,
    \label{eq:SG_identity}
\end{equation}
also found using the propagators' defining identities in Eqs.~\eqref{eq:orthonormal} and~\eqref{eq:complete}. With the appearance of the Heaviside theta functions we must be cautious. We go back to the point split form of the partition function, namely $[\theta(z_0)+\theta(-z_0)]\langle\textrm{in} |\textrm{in}\rangle$, where we break about the function into $\pm z_0$ components. In such a case notice too that the causal function, $\sprop$, in $[\sprop \Delta]$ gets split, and for $\theta(z_0)(\theta(-z_0))$ arguments, only the $\langle\psi(x)\bar{\psi}(y)\rangle(\langle \bar{\psi}(y)\psi(x)\rangle)$ is projected. This is important for then when we look at products of $[\sprop\Delta]^n$ containing in Eq.~\eqref{eq:SG_identity}, for each $\theta(\pm z_0)$ half contraction with either $\langle\psi(x)\psi^{\dagger}(x')\rrangle$ or $\langle\psi^{\dagger}(x')\psi(x)\rrangle$ will act as an identity element. Alternatively, one may add to Eq.~\eqref{eq:SG_identity} the following factor:  $\theta(z_0)\langle\psi^{\dagger}(x')\psi(x)\rrangle+\theta(-z_0)\langle\psi(x)\psi^{\dagger}(x')\rrangle$, which is zero when placed in the partition function. What this accomplishes is that we may then write for the partition function
\begin{equation}
    \langle\text{in}|\text{in}\rangle 
    =|c_{v}|^{2}\exp\Bigl\{-\sum_{n=1}^{\infty}\frac{1}{n}\mathrm{Tr}_3[\sprop\gamma_0 \spropbar \gamma_0 +I]^{n}\Bigr\}=|c_v|^2\mathrm{Det}_3[-\sprop\gamma_0 \spropbar \gamma_0]\,.
    \label{eq:inin_3}
\end{equation}
We can immediately see the resemblance in this form to the imaginary part of the effective action. The above analysis is important to correctly reproduce the identity element; naively taking the coincident limit in Eq.~\eqref{eq:SG_identity} would lead to an erroneous factor of two, troublesome to identify without splitting the partition function.

At this point for the purpose of comparison let us digress to make the connection to the imaginary part of the effective action. And in so doing we will also extend Eq.~\eqref{eq:inin_3} to a (3+1)-dimensional determinant and trace. First, since the functional determinant is regulated for $c_v=\mathrm{Det}_4[i\slashed{D}-m+i\epsilon]$ with free-field subtraction\footnote{Since $\int^\infty_0ds\,s^{\varepsilon-1}e^{-i(\hat{H}-i\epsilon)s}=\varepsilon^{-1} -\gamma-\ln(\hat{H}-i\epsilon)-i\pi/2+\mathcal{O}(\varepsilon)$, for example using dimensional regularization, to put the logarithm into worldline form a free-field subtraction, or other suitable regularization, is necessary to obtain meaningful results.} as
\begin{equation}
    \mathrm{Det}_4\Bigl[\frac{i\slashed{D}-m+i\epsilon}{i\slashed{\partial}-m+i\epsilon}\Bigr]
    =\mathrm{Det}_4\Bigl[\frac{i\slashed{D}-m+i\epsilon}{i\slashed{\partial}-m+i\epsilon}\frac{\Lambda}{\Lambda}\Bigr]\,,
\end{equation}
at this point too let us regulate the determinant as it appears above in~\eqref{eq:inin_3} so as to include a free-field subtraction. Let us treat this subtraction as implicit in the definition of the determinant, and hence also the probability for $N$-pair production, for calculations to come for brevity. We have furthermore introduced a dimensionfull constant factor, $\Lambda$, that will later be useful. Next, taking the complex conjugate of $c_v$ one can find, combining the determinants that
\begin{equation} |c_v|^2\mathrm{Det}_4[\Lambda^{-2}]=\mathrm{Det}_4\bigl[(i\slashed{D}-m+i\epsilon)\gamma_0(i\slashed{D}-m-i\epsilon)\gamma_0\Lambda^{-2}\bigr]\,.
    \label{eq:imag_partition}
\end{equation}
Now we will show that the inverse of the above indeed satisfies Eq.~\eqref{eq:inin_3}. Let us extend Eq.~\eqref{eq:inin_3} to a functional determinant over (3+1)-dimension. First notice that since any time may be used for the special case of the in-in partition function, or the SK contour with no operator insertions,\footnote{The same identification is not possible for objects with operator insertions such as is the case for the in-in propagator; there we required at least that all common times $x_0$ and $y_0$ be greater than the times of the inserted Dirac operators to maintain time-ordering.} one may equally well average over all possible times such that for Eq.~\eqref{eq:in-in_gen} we have
\begin{equation}
    \int d^{3}xd^{3}y\bar{\psi}(x)\Delta(x,y)\psi(y)=\Lambda^{2}\int d^{4}xd^{4}y\bar{\psi}(x)\Delta(x,y)\psi(y)\,,\label{eq:time_measure}
\end{equation}
with $\Lambda$ representing a measure, and with dimension, of inverse time to later be determined. The above identification is important in that with it, one may assume a complete set of states in time such that $I_{x_0}=\int dx_{0}|x_{0}\rangle\langle x_{0}|$ permitting an integration by parts in time, in addition to those already in space. This distinction may also permit us to write the functional form of the propagators after using $\int d^4x|x\rangle\langle x|=I$. Orders of $\Lambda^2$ also help us to keep track of the powers of $S^c\Delta$ that indicate pair production probability. We can find that Eq.~\eqref{eq:inin_3} becomes
\begin{equation}
    \langle\text{in}|\text{in}\rangle 
    =|c_v|^2\mathrm{Det}_4[\Lambda^{-2}]\mathrm{Det}_4[-\sprop\gamma_0 \spropbar \gamma_0\Lambda^{2}]\,,
    \label{eq:inin_4}
\end{equation}
indeed satisfying Eq.~\eqref{eq:imag_partition}. Here now
\begin{equation}
    \mathrm{Tr}_4[\sprop\gamma_0 \spropbar \gamma_0]^{n}\coloneqq\prod_{i=1}^{n}\int d^{4}x_{i}d^{4}y_{i}\mathrm{tr}[\sprop(x_{1},y_{1})\gamma_0\spropbar(y_{1},x_{2})\gamma_0\,...\,\sprop(x_{n},y_{n})\gamma_0\spropbar(y_{n},x_{1})\gamma_0]\,,
\end{equation}
which also defines $\mathrm{Det}_4$. 

The utility of expressing the $\sprop\Delta$ factors into the form of Eq.~\eqref{eq:inin_3} or Eq.~\eqref{eq:inin_4}, is that we can apply the property of the determinant to factor out the $\gamma_0$'s and introduce $\gamma_5$'s--as is done to arrive at a quadratic form of the Dirac operator for the effective action.
\begin{equation}
    \langle\text{in}|\text{in}\rangle 
    =|c_v|^2\mathrm{Det}_4[\Lambda^{-2}]\mathrm{Det}_4[-\sprop \spropbar \Lambda^{2}]
    =|c_v|^2\mathrm{Det}_4[\Lambda^{-2}]\mathrm{Det}_4[-\sprop\gamma_5 \spropbar \gamma_5\Lambda^{2}]\,.
    \label{eq:gamma_5_replace}
\end{equation}
We have a more convenient object as the arguments of our Bell polynomials now. The arguments of our Bell polynomials can now be written as, after integrating out the spacetime d.o.f.,
\begin{equation}
    p_n\coloneqq -\mathrm{Tr}_4\Bigl[\frac{-1}{i\hat{\slashed{D}}-m+i\epsilon}\gamma_{5}\frac{1}{i\hat{\slashed{D}}-m-i\epsilon}\gamma_{5}\Lambda^2 +I \Bigr]^n\,,
    \label{eq:p_n1}
\end{equation}
where the (3+1)-dimensional trace acts as usual: $\mathrm{Tr}_4\mathcal{O}=\int d^4 x \langle x|\mathcal{O}|x\rangle$. We can find a simplified expression for the $n$-th order element by noting that
\begin{equation}
    \partial_{m^{2}} \Bigl(\frac{1}{i\hat{\slashed{D}}-m+i\epsilon}\gamma_{5}\frac{1}{i\hat{\slashed{D}}-m-i\epsilon}\gamma_{5}\Bigr)=-\Bigl[\frac{1}{i\hat{\slashed{D}}-m+i\epsilon}\gamma_{5}\frac{1}{i\hat{\slashed{D}}-m-i\epsilon}\gamma_{5}\Bigr]^{2}\,.
\end{equation}
And hence we can determine more generally for $n\geq 1$ that
\begin{equation}
    \Bigl[\frac{-1}{i\hat{\slashed{D}}-m+i\epsilon}\gamma_{5}\frac{1}{i\hat{\slashed{D}}-m-i\epsilon}\gamma_{5}\Bigr]^{n}=\frac{1}{(n-1)!}\partial_{m^{2}}^{n-1}\Bigl(\frac{-1}{i\hat{\slashed{D}}-m+i\epsilon}\gamma_{5}\frac{1}{i\hat{\slashed{D}}-m-i\epsilon}\gamma_{5}\Bigr)\,.
\end{equation}
Finally, we can then find the simplified expression for the operator in~\eqref{eq:p_n1},
\begin{align}
    &\Bigl[\frac{-1}{i\hat{\slashed{D}}-m+i\epsilon}\gamma_{5}\frac{1}{i\hat{\slashed{D}}-m-i\epsilon}\gamma_{5}\Lambda^2+I \Bigr]^n\notag
    \\=&I+\Lambda^2L_{n-1}^{(1)}(-\Lambda^{2}\partial_{m^2})\Bigl(\frac{-1}{i\hat{\slashed{D}}-m+i\epsilon}\gamma_{5}\frac{1}{i\hat{\slashed{D}}-m-i\epsilon}\gamma_{5}\Bigr)\,,
    \label{eq:I_divergent}
\end{align}
where the associated Laguerre polynomial of parameter one is
\begin{equation}
    L_{n-1}^{(1)}(-\Lambda^2\partial_{m^2})=\sum_{k=1}^{n}\binom{n}{k}\frac{1}{(k-1)!}(\Lambda^2\partial_{m^{2}})^{k-1}\,.
\end{equation}
The $I$ as written in Eq.~\eqref{eq:I_divergent} once summed over to infinity represent an anticipated divergent field-free contribution that according to our regularization scheme drops out. Now let us evaluate the following operator:
\begin{align}
    \langle x|\frac{1}{i\hat{\slashed{D}}-m+i\epsilon}\gamma_{5}\frac{1}{i\hat{\slashed{D}}-m-i\epsilon}\gamma_{5}|y\rangle &=\frac{1}{2}\langle x|\frac{1}{\hat{\slashed{D}}^{2}+m^{2}-i\epsilon}+\frac{1}{\hat{\slashed{D}}^{2}+m^{2}+i\epsilon}|y\rangle+\mathcal{O}(\epsilon)\nonumber\\&=\frac{1}{2}\Bigl\{\int_{\Gamma}ds+\int_{h}ds\Bigr\} g(x,y,s)\,.
\end{align}
We can see the appearance of the kernel of the $\rho_h$ function as anticipated. Further the semi-circle contour over the origin, $\Gamma$, vanishes for $x_0=y_0$.

We finally find for the arguments of the Bell polynomials the following compact form: 
\begin{equation}
    p_n=\Lambda^2\frac{1}{2}L_{n-1}^{(1)}(-\Lambda^{2}\partial_{m^2})\mathrm{tr}\int d^4x\int_{h}ds\,g(x,x,s)\,,
    \label{eq:p_n_final}
\end{equation}
where now the probability for $N$ pair creation reads
\begin{equation}
    P_N = \frac{|c_v|^2}{N!}B_N\Bigl(  p_1,...,(N-1)!\,p_N \Bigr)\,.
    \label{eq:P_N_final}
\end{equation}
We emphasize that the above formulation holds for \textit{any} background field in QED. And, with the above formulation evaluation of pair production for any $N$ pairs is now no more difficult than evaluating the vacuum non-persistence. One need only determine, for example for fields that only possess simple poles in the complex $s$ plane, the location of such poles.

Having put the resummed structure of the $N$-pair creation probability into a convenient worldline form--one that closely resembles the imaginary part of the effective action, we can now use a matching to the effective action to fix the factor $\Lambda^2$. Fixing follows from the physical demand that a leading order contribution to the effective action match the leading order contribution to the probability of creating a single pair. We can cast this statement into a more concrete form through $\lim_{m^2\to\infty}\ln\langle \text{in}|\text{in}\rangle=0$. What this accomplishes is to pick out the dominant singularity, or worldline instanton for that matter. Thus only $p_1$ over the sum over all $p_n$ will contribute. Then, for example, for background fields that admit discrete instantons, or removable singularities in propertime, only the lowest order pole would contribute. Since $p_1$ carries a sole factor of $\Lambda^2$ from $p_1=(\Lambda^2/2)\mathrm{tr}\int d^4x\int_{h}ds\,g(x,x,s)$ and likewise the imaginary part of the effective action from Eq.~\eqref{eq:ImGamma} indicates that the only difference (apart from the minus sign) resides in an additional factor of $s^{-1}$, we can see that $\Lambda^{-2}$ can be identified with the dominant pole contribution. Let us put this into a more precise form, which resembles a ground state extraction, of
\begin{equation}
    \Lambda^{-2} = \lim_{m^2\to\infty}\frac{1}{\mathrm{Im}\Gamma}\partial_{m^2}\mathrm{Im}\Gamma\,.
\end{equation}
This is, for example, for the case of a homogeneous electric field the location in imaginary propertime of the dominant pole, or $\Lambda^2=eE/\pi$.

Let us confirm Eq.~\eqref{eq:P_N_final} does still reproduce the vacuum non-persistence once summed over to infinity. To do so let us write the sum over all $P_N$ in exponential form as
\begin{equation}
    \sum_{N=0}^\infty P_N=|c_v|^2\exp\Bigl[ \sum_{n=1}^\infty \frac{p_n}{n} \Bigr]\,.
    \label{P_N_sum}
\end{equation}
Now as $p_n$ is written in Eq.~\eqref{eq:p_n_final} one may write for the $n$ dependent terms contained in the Laguerre polynomial 
\begin{equation}
    L_{n-1}^{(1)}(-\Lambda^2\partial_{m^2})\to L_{n-1}^{(1)}(i\Lambda^2s)\,.
\end{equation}
Then we can write for the sum over the polynomials
\begin{equation}
    \sum_{n=1}^\infty \frac{1}{n}L_{n-1}^{(1)}(i\Lambda^2s) = \int^1_0 du\,\frac{1}{(1-u)^2}\exp\Bigl(\frac{-i\Lambda^2su}{1-u} \Bigr)\,,
\end{equation}
where we have made use of the generating function of associated Laguerre polynomials $\sum_{n=0}^\infty L_n^{(1)}(u)x^m=(1-x)^{-2}\exp(-ux/(1-u))$. Then we perform the change of variables to $v=u/(1-u)$ where we can find
\begin{equation}
    \sum_{n=1}^\infty \frac{1}{n}L_{n-1}^{(1)}(i\Lambda^2s) = \int^\infty_0 dv\, \exp(-i\Lambda^2sv)=\frac{1}{i\Lambda^2s}\,,
\end{equation}
where we have assumed $\mathrm{Im}(s)<0$. Placing the above into Eq.~\eqref{P_N_sum} we find
\begin{equation}
    \sum_{N=0}^\infty P_N=|c_n|^2\exp\Bigl[ -\frac{i}{2}\mathrm{tr}\int d^4x\int_{h}\frac{ds}{s}\,g(x,x,s) \Bigr]\,.
\end{equation}
One can then see from Eq.~\eqref{eq:ImGamma} that indeed the sum over all $N-$pair creation does reproduce the imaginary part of the effective action.

Having confirmed the expression for $N-$pair creation probability, in Eqs.~\eqref{eq:p_n_final} and~\eqref{eq:P_N_final}, let us consider a concrete example; we use the background of homogeneous and parallel electric and magnetic fields with $\boldsymbol{E}=E\hat{x}^{3}$ and $\boldsymbol{B}=B\hat{x}^{3}$. The propagator, and kernel, are exactly known in a homogeneous electromagnetic field~\cite{Schwinger:1951nm}, and in a Fock-Schwinger gauge the kernel reads
\begin{align}
g(x,y,s) & =\frac{e^{2}EB}{(4\pi)^{2}}\exp[-im^{2}s+i\varphi(x,y,s)]\sin^{-1}(eBs)\sinh^{-1}(eEs)\,\Phi\,,\\
\varphi(x,y,s) & =\frac{1}{2}x_{\mu}eF_{\;\nu}^{\mu}y^{\nu}+\frac{1}{4}\bigl[(z_{3}^{2}-z_{0}^{2})eE\coth(eEs)+(z_{1}^{2}+z_{2}^{2})eB\cot(eBs)\bigr]\,,\\
\Phi & =[\cos(eBs)+i\sin(eBs)\sigma^{12}]\times[\cosh(sEs)+\sinh(eEs)\gamma_{5}\sigma^{12}]\,.
\end{align}
Determination of the contour integrals over the poles is straightforward, and we find for the probability to create a single pair as
\begin{align}
    P_{1} &=|c_v|^2p_1 =|c_v|^2\frac{\Lambda^{2}}{2}\int d^{4}x\,\mathrm{tr}\sum_{n=1}^{\infty}\ointclockwise_{-i\frac{n\pi}{eE}}ds\,g(x,x,s)\\
    &=|c_v|^2\frac{e^{2}EB\,VT}{4\pi^{2}}\sum\limits _{n=1}^{\infty}\coth\Bigl(\frac{n\pi B}{E}\Bigr)\exp\Bigl(-\frac{n\pi m^{2}}{eE}\Bigr)\,.\label{eq:aomega}
\end{align}
This agrees with the exact result found by determining the Bogoliubov coefficients in Ref.~\cite{fradkin1991quantum}. Using Eq.~\eqref{eq:p_n_final} we can further determine the probability for $N$-pair creation as~\eqref{eq:P_N_final} with
\begin{equation}
    p_n=\frac{e^{2}EB\,VT}{4\pi^{2}}\sum\limits _{n'=1}^{\infty}L_{n-1}^{(1)}(n')\coth\Bigl(\frac{n'\pi B}{E}\Bigr)\exp\Bigl(-\frac{n'\pi m^{2}}{eE}\Bigr)\,.
\end{equation}
Pair production probabilities in homogeneous fields have also been examined using a Bogoliubov coefficient formalism~\cite{Nikishov:1969tt}, wherein one first determines the eigendecomposition of the Dirac equation and then sums over the eigenspectrum; amplitudes can then be found from truncation. A distinct motivation of our approach is that one does not need to determine the eigenspectrum to calculate quantities related to Schwinger pair production.

Let us next consider another example, one that does not rely on determining poles in imaginary propertime; this is the case of a Sauter background field~\cite{Sauter:1931zz}. One merit of casting the $N-$creation pair probability in the worldline formalism is the ability to use the worldline instanton method~\cite{Affleck:1981bma,Dunne:2005sx}, which is capable of treating complicated fields without knowledge of the Dirac equation's eigenspectrum in the background field. And the Sauter field provides a good showcase for the technique; we refer the reader to Ref.~\cite{Dunne:2005sx,Dunne:2006st} for further details on the approach. The background field here is $E(x_0)=E\,\mathrm{sech}^2(\omega x_0)$ with gauge field $A_3(x_0)=-(E/\omega)\,\mathrm{tanh}(\omega x_0)$. The worldline instanton technique is a semi-classical evaluation; thus let us first determine the classical solutions. The kernel, $g(x,x,s)$, here is given in its worldine path integral representation, Eq.~\eqref{eq:worldline_pi}; however let us first do the convenient transformation $\tau\to T\tau$, then we find that
\begin{equation}
    g(x,y,s)=i\int\mathcal{D}x\,e^{i\int_{0}^{1}d\tau[-m^2s-\frac{\dot{x}^{2}}{4s}-eA\cdot \dot{x}]}\mathcal{P}e^{-\frac{is}{2}\int_{0}^{1}d\tau\, eF\cdot\sigma}\,.
\end{equation}
Then equations of motion are just the Lorentz force equation, $\ddot{x}_\mu=2seF_{\mu\nu}\dot{x}^\nu$ and the mass-shell condition in Schwinger propertime: $m^2=\dot{x}^2/(4s^2)$ coming from the $s$ integral. $\dot{x}_1=\dot{x}_2=0$ because of periodicity. Note that the spin factor does not contribute to stationary points since for the background electric field in Minkowski space it is imaginary in the worldline action. Next, using the fact that $\dot{x}_0^2-\dot{x}_3^2=4m^2s^2$ and that $\dot{x}_3=-2seA_3$, we need only evaluate
\begin{equation}
    \dot{x}_0(\tau)=\pm 2ms\sqrt{\gamma^{-2}\,\mathrm{tanh}^2[\omega x_0(\tau)]+1}\,,
\end{equation}
with Keldysh adiabaticity parameter, $\gamma=m\omega/(eE)$~\cite{Keldysh:1965ojf}. It has the following solution
\begin{equation}
    x_0(\tau)=\frac{1}{\omega}\mathrm{arcsinh}\Bigl[ \frac{\gamma}{\sqrt{1+\gamma^2}}\sinh \Bigl( 2\omega ms\tau \frac{\sqrt{1+\gamma^2}}{\gamma} \Bigr)  \Bigr]\,.
\end{equation}
Periodicity requirements $(x(0)=x(1))$ dictate that stationary points in propertime are at $s_n=-in\pi \gamma/(\omega m\sqrt{1+\gamma^2})\; \forall n\in \mathbb{Z}^+$. Likewise one may determine from $\dot{x}_3=-2seA_3$ that
\begin{equation}
    x_3(\tau)= \frac{1}{\omega\sqrt{1+\gamma^2}}\mathrm{arcsinh}\Bigl[ \gamma\cosh\Bigl( 2\omega ms\tau \frac{\sqrt{1+\gamma^2}}{\gamma}  \Bigr)  \Bigr]\,.
\end{equation}
Now the bosonic part of the worldline action can readily be evaluated on the instantons to find the dominant exponential suppression 
\begin{equation}
    -im^{2}s-i\frac{\dot{x}^{2}}{4s}-ie\int_{0}^{1}d\tau\,A\cdot\dot{x}=-\frac{n\pi m^{2}}{eE}\frac{2}{1+\sqrt{1+\gamma^{2}}}\,.
    \label{eq:exp_factor}
\end{equation}
In the $\gamma\to0$ limit the factor for the homogeneous field is reproduced, and in the $\gamma\to\infty$ limit the perturbative regime appears. 
The spin factor evaluated about the classical instanton path simplifies 
\begin{equation}
    \mathrm{tr}\mathcal{P}e^{-\frac{is}{2}\int_{0}^{1}d\tau\, eF\cdot\sigma}=4\cos\Bigl[iseE\int_{0}^{1}d\tau\,\mathrm{sech}^{2}(\omega x_{0})\Bigr]=4\cos(n\pi)\,.
\end{equation}
Finally, since pair production is dominated about the exponential factor, let us treat the weakly inhomogeneous case in which $\gamma$ is small for the evaluation of fluctuation determinant. Expanding about $x(\tau)=x_{cl}(\tau)+\eta(\tau)$, where $x_{cl}(\tau)$ is the classical solution above, and $s=s_n+u$ we find the fluctuation determinant and fluctuations in propertime reduce to
\begin{equation}
    2\int_{i\epsilon}^{-i\epsilon}\!du\,e^{i\frac{\dot{x}_{cl}^{2}}{4s_{n}^{3}}u^{2}}\mathrm{Det}^{-1/2}\Bigl[\frac{d^2}{d\tau^2}\frac{1}{4s}\eta_{\mu\nu}-\frac{e}{2}F_{\mu\nu}\frac{d}{d\tau}+\frac{e}{2}\partial_{\mu}F_{\sigma\nu}\dot{x}^{\sigma}\Bigr]\approx\frac{-2\pi}{(4\pi)^2s_n\cos(n\pi)}\,,
\end{equation} 
where we have used that $F_{\mu\nu}\approx - E(\delta_{\mu}^0\delta_\nu^3-\delta_{\mu}^3\delta_\nu^0)$ in the determinant. The full fluctuation determinant is treated in Ref.~\cite{Dunne:2006st}, but is beyond the scope of our analysis here. $\Lambda^{2}=eE(1+\sqrt{1+\gamma^2})/(2\pi)$. Gathering everything together we may find for the probability for a single pair to be produced for weakly inhomogeneous fields as
\begin{equation}
    P_{1} \approx |c_v|^2\frac{e^2E^2\sqrt{1+\gamma^2}(1+\sqrt{1+\gamma^2})VT}{8\pi^3}\,\sum_{n=1}^{\infty}\frac{1}{n}\exp\Bigl(-\frac{n\pi m^{2}}{eE}\frac{2}{1+\sqrt{1+\gamma^{2}}}\Bigr)\,.
\end{equation}
From the above one may then quickly discover the arguments determining the $N-$pair production probability (given in Eq.~\eqref{eq:p_n_final}) as
\begin{equation}
    p_n \approx \frac{e^2E^2\sqrt{1+\gamma^2}(1+\sqrt{1+\gamma^2})VT}{8\pi^3}\,\sum_{n'=1}^{\infty}L^{(1)}_{n-1}(n')\frac{1}{n'}\exp\Bigl(-\frac{n'\pi m^{2}}{eE}\frac{2}{1+\sqrt{1+\gamma^{2}}}\Bigr)\,.
\end{equation}
A merit of our construction is that only the imaginary stationary points in Schwinger propertime contribute. In the above we have taken that $\int d^4x=VT$ in accordance with our weak field approximation.

We have illustrated two ways one may use Eq.~\eqref{eq:p_n_final} to determine the $N-$pair creation probability. However, there are a myriad of different approaches as one's disposal; let us name a few: In addition to worldline instanton/semi-classical techniques~\cite{Affleck:1981bma,Dunne:2005sx,Dunne:2006st,Dumlu:2011cc,Ilderton:2015lsa,Copinger:2020feb} and exact kernel constructions~\cite{Schwinger:1951nm,Dunne:2004nc}, perturbative technqiues are a mainstay approach in the worldline formalism~\cite{ChrisRev}; for a non-perturbative picture essential for the Schwinger effect one may expand about analytically tractable profiles. Heat kernel techniques are also well-studied~\cite{Vassilevich:2003xt,Fliegner:1994zc,Bastianelli:2008vh,Fecit:2025kqb}. One may also apply Borel resummation schemes~\cite{Dunne:2021acr}. And numerical technquies are also widely used including Monte-Carlo based computations~\cite{Gies:2001tj,Gies:2001zp,Ahumada:2025poa} and numerical semi-classical techniques~\cite{Schneider:2018huk}.

Having seen how the in-in partition function leads to an in-in worldline description for the $N-$pair creation probability, let us now look at the in-in causal propagator. This is from Eq.~\eqref{eq:S_in^c_final}
\begin{equation}
    \Sprop(x',y')=i|c_{v}|^{2}\langle \mathcal{T}e^{i\int d^{3}xd^{3}y\bar{\psi}(y)\Delta(y,x)\psi(x)}\psi(x')\bar{\psi}(y)\rrangle\,,
\end{equation}
where 
\begin{equation}
    x_0,y_0>x_0',y_0'\,
    \label{eq:xy_arg}
\end{equation} 
One may take equally well take $x_0,y_0\to\infty$. The key difference between the in-in propagator and the in-in partition function is indeed the arguments for $x_0$ and $y_0$. Whereas for the in-in partition function any value may be applied, here we have Eq.~\eqref{eq:xy_arg}, and hence the major simplification shown above may no be applied to the resummation of the in-in propagator. The affected simplifications include both the averaging over all times as represented in Eq.~\eqref{eq:time_measure}, and also the properties of the determinant, in Eq.~\eqref{eq:gamma_5_replace}, leading ultimately to $\gamma_0\to\gamma_5$ in the $\sprop\Delta$ expressions. However, we may still present concrete in-in worldline forms. Therefore, let us repeat the same steps as employed above for the in-in partition function to find
\begin{align}
    \Sprop(x',y') & =|c_{v}|^{2}\Bigl\{\sum_{n=0}^{\infty}\frac{(-i)^{n}}{n!}\langle\mathcal{T}[-i\psi_{x}\bar{\psi}_{y}\Delta_{yx}]^{n}\rrangle\sprop_{x'y'}\\
     & \quad+\sum_{n=1}^{\infty}\frac{(-i)^{n-1}n}{n!}\langle\mathcal{T}[-i\psi_{x}\bar{\psi}_{y}\Delta_{yx}]^{n-1}\rrangle\sprop_{x'y}\Delta_{yx}\sprop_{xy'}\notag\\
     & \quad+\sum_{n=2}^{\infty}\frac{(-i)^{n-2}n(n-1)}{n!}\langle\mathcal{T}[-i\psi_{x}\bar{\psi}_{y}\Delta_{yx}]^{n-2}\rrangle\sprop_{x'y_{1}}\Delta_{y_{1}x_{1}}\sprop_{x_{1}y_{2}}\Delta_{y_{2}x_{2}}\sprop_{x_{2}y'}\Bigr\}+...\nonumber\\
     & =\sprop_{x'y'}+\sprop_{x'y}\Delta_{yx}\sprop_{xy'}+\sprop_{x'y_{1}}\Delta_{y_{1}x_{1}}\sprop_{x_{1}y_{2}}\Delta_{y_{2}x_{2}}\sprop_{x_{2}y'}+...\,,
\end{align}
where as before we have that $x^0_i=x^0$ and $y^0_i=y^0$ for all $i$. We can see that the addition of the interaction term has led to a geometric series definition for the in-in causal propagator. Here the sum of all such diagrams in the Dyson series modifies the in-out propagator by a self-energy expression. One worldline representation for the propagators in the series follows from Eq.~\eqref{eq:SDelta_1} for each pair of $\sprop\Delta$. Likewise, using Eq.~\eqref{eq:SS_cancel}, one may equally well write the simplified expression that uses Eq.~\eqref{eq:SDelta_2}, which extracts the poles in the imaginary complex plane. Last, we note that we can formally write for the series
\begin{equation}
    \Sprop(x',y')=\Bigl[\frac{I_3}{I_3-\sprop\Delta}\sprop\Bigr]_{x',y'}\,,
\end{equation}
in analogy to the description used for the functional determinant in Eq.~\eqref{eq:det_3}. Resummation of the geometric series for the in-in propagator will require new techniques then those used for the in-in partition function, and therefore we leave this problem as one for future work that will be reported elsewhere. However, at this point let us remark in analogy to truncating the imaginary part of the effective action to leading orders in pair production, here too for the in-in partition function one may well approximate
\begin{equation}
    \Sprop(x',y') \approx\sprop_{x'y'}+\sprop_{x'y}\Delta_{yx}\sprop_{xy'}\,,
\end{equation}
which would be valid description to include the effects of up to the creation of a \textit{single} pair of particles, c.f., $P_1$ for the single pair production probability.

%%%%%%%%%%%%%%%%%%%%%%%%%%%%%%%%%%%%%%%%%%%%%%%%%%%%%%%%%%%%%%%%%%%%%%%
\section{Conclusions}
\label{sec:conclusions}

In this work, we have derived a real-time framework, specifically the in-in formalism, for pair production to all orders in strong-field QED via two approaches: both from Bogoliubov coefficients and from the SK closed-time path. By mapping the Bogoliubov coefficients from in-out transition amplitudes to the results of the SK closed-time path formalism, we have bridged the gap between the well-established in-out transition amplitudes  and the physically relevant in-in expectation values required for real-time dynamics. A first-quantized form is manifest from the in-out ingredients, thus also furnishing an in-in worldline formalism. Both approaches yield a common augmentation to the in-out partition function and propagator in the form of an exponential non-local term (see, e.g. Eq.~\eqref{eq:S_in^c_final}), which captures the various singularities in imaginary Schwinger proper time associated with the Schwinger effect. Overall, this work significantly expands the scope of the worldline formalism and establishes a direct connection to the SK closed-time path framework.

Interestingly, we notice that for the partition function, because no operators are present beyond this non-local term, any time, not only the asymptotic `out' time, may be used, greatly simplifying the overall structure. This allows for a resummation and a more compact, first-quantized expression for the $N$-pair production probability, $P_N$ in Eq.~\eqref{eq:P_N_final}. However, such simplifications do not extend to the in-in propagator, and resumming the contributions for all $N$ pairs remains an open problem. To advance this direction, a more rigorous analysis of the asymptotic characteristics of the first-quantized in-out propagator—for arbitrary fields and to all orders—would be highly beneficial. We suggest that an investigation of worldline instantons on the open line~\cite{DegliEsposti:2022yqw,DegliEsposti:2024rjw} represents a vital first step.

While this work focuses exclusively on strong-field QED, the formalism is, in principle, applicable to the scalar case as well. One would only need to perform the following replacements in, for example, in Eq.~\eqref{eq:S_in^c_final}: $\gamma_0\to\overleftrightarrow{D}_0$, while removing both the spin factor, $\mathcal{P}\exp(-(i/2)\int^T_0d\tau\,\sigma\cdot eF)$, and the leading $i\slashed{D}_x+m$  factor from the propagators. Furthermore, extensions to non-Abelian gauge fields and curved spacetime backgrounds are feasible, as the Bogoliubov coefficient machinery utilized in Sec.~\ref{sec:in-in} has already been developed for these contexts~\cite{Buchbinder:1981hu}. Such extensions represent compelling avenues for future research.

\begin{acknowledgments}
We are grateful to Kenji Fukushima and Di-Lun Yang for fruitful discussions in the earlier stages of this work, that have led to its betterment. PC would also like to thank James P. Edwards and Karthik Rajeev, also for their insightful discussions. 
This work is supported in part by the National Key Research
and Development Program of China under Contract No. 2022YFA1605500,
by the Chinese Academy of Sciences (CAS) under Grant No. YSBR-088 and by National Natural Science Foundation of China (NSFC) under Grants No.~12135011 (SP).
This work is supported in part by the EPSRC Standard Grant  EP/X02413X/1 (PC), and by the WPI program ``Sustainability with Knotted Chiral Meta Matter (WPI-SKCM${}^2$)" at Hiroshima University (PC). PC would further like to acknowledge support from the Research Start-up Support Fund of WPI-SKCM$^2$. \\ 
\emph{Note added:} Recently, we were informed of Ref.~\cite{Fukushima:2025eyt} which works on a similar topic and appeared on arXiv on the same day.
\end{acknowledgments}

\appendix

\bibliography{references}
\bibliographystyle{JHEP}

\end{document}